\documentclass[fleqn,usenatbib]{mnras}

\usepackage{newtxtext,newtxmath}
\usepackage[T1]{fontenc}
\usepackage{booktabs}

\usepackage{graphicx} % For including graphics
\usepackage{caption}
\usepackage{subcaption} % For creating subfigures
\usepackage{lipsum}
\usepackage{float}

\usepackage{xcolor}

\newcommand{\im}{\text{im}}
\renewcommand{\sp}{\text{sp}}

\usepackage{lipsum} % For placeholder text

\newcommand{\numberfootnote}[1]{%
    \begingroup
    \renewcommand{\thefootnote}{\textsuperscript{\arabic{footnote}}}%
    \footnote{#1}%
    \endgroup
}

\DeclareRobustCommand{\VAN}[3]{#2}
\let\VANthebibliography\thebibliography
\def\thebibliography{\DeclareRobustCommand{\VAN}[3]{##3}\VANthebibliography}

\usepackage{graphicx}	% Including figure files
\usepackage{amsmath}	% Advanced maths commands
\usepackage{titlesec}
\usepackage{subcaption}

\title[AstroCLIP]{AstroCLIP: A Cross-Modal Foundation Model for Galaxies}

\author[Liam Parker]{
Liam Parker,$^{1}$
\thanks{E-mail:lparker@flatironinstitute.org}
\thanks{Present Address: 162 5th Ave, New York, NY 10010, USA}
Francois Lanusse,$^{1,3}$
Siavash Golkar,$^{1}$
Leopoldo Sarra,$^{1}$
Miles Cranmer,$^{4}$
\newauthor{Alberto Bietti,$^{1}$ Michael Eickenberg,$^{1}$  Geraud Krawezik,$^{1}$ Michael McCabe,$^{1, 5}$ Rudy Morel,$^{1}$ Ruben Ohana,$^{1}$}
\newauthor{Mariel Pettee,$^{1,6}$ Bruno Régaldo-Saint Blancard,$^{1}$ Kyunghyun Cho,$^{7, 8, 9}$ Shirley Ho$^{1, 7, 10}$ and}
\newauthor{The Polymathic AI Collaboration}
\\
$^{1}$The Flatiron Institute, 162 5th Ave, New York, NY, 10010, USA\\
$^{2}$Université Paris-Saclay, Université Paris Cité, CEA, CNRS, AIM, Paris, 91190, France\\
$^{3}$Department of Astronomy, University of Cambridge, Madingley Rd, Cambridge, CB3 0HA, UK\\
$^{4}$Department of Computer Science, University of Colorado, Boulder, 430 UCB, 1111 Engineering Dr, Boulder, CO, 80309, USA\\
$^{5}$Lawrence Berkeley National Laboratory, Berkeley, 1 Cyclotron Rd, CA, 94720, USA\\
$^{6}$Center for Data Science, New York University, 60 5th Ave, New York, NY, 10011, USA\\
$^{7}$Prescient Design, Genentech, 149 5th Ave, New York, NY, 10010, USA\\
$^{8}$CIFAR Learning in Machines and Brains Fellow, Toronto, ON M5G 1M1, Canada \\
$^{9}$Department of Astrophysics, Princeton University, 4 Ivy Lane, Princeton, NJ, 08544, USA
}

\date{Accepted XXX. Received YYY; in original form ZZZ}

\pubyear{2024}

\begin{document}
    
\label{firstpage}
\pagerange{\pageref{firstpage}--\pageref{lastpage}}
\maketitle

\begin{abstract}
We present AstroCLIP, a single, versatile model that can embed both galaxy images and spectra into a shared, physically meaningful latent space. These embeddings can then be used - without any model fine-tuning - for a variety of downstream tasks including (1) accurate in-modality and cross-modality semantic similarity search, (2) photometric redshift estimation, (3) galaxy property estimation from both images and spectra, and (4) morphology classification. Our approach to implementing AstroCLIP consists of two parts. First, we embed galaxy images and spectra separately by pretraining separate transformer-based image and spectrum encoders in self-supervised settings. We then align the encoders using a contrastive loss. We apply our method to spectra from the Dark Energy Spectroscopic Instrument and images from its corresponding Legacy Imaging Survey. Overall, we find remarkable performance on all downstream tasks, even relative to supervised baselines. For example, for a task like photometric redshift prediction, we find similar performance to a specifically-trained ResNet18, and for additional tasks like physical property estimation (stellar mass, age, metallicity, and sSFR), we beat this supervised baseline by 19\% in terms of $R^2$. We also compare our results to a state-of-the-art self-supervised single-modal model for galaxy images, and find that our approach outperforms this benchmark by roughly a factor of two on photometric redshift estimation and physical property prediction in terms of $R^2$, while remaining roughly in-line in terms of morphology classification. Ultimately, our approach represents the first cross-modal self-supervised model for galaxies, and the first self-supervised transformer-based architectures for galaxy images and spectra. 

\end{abstract}


%%%%%%%%%%%%%%%%%%%%%%%%%%%%%%%%%%%%%%%%%%%%%%%%%%

%%%%%%%%%%%%%%%%% BODY OF PAPER %%%%%%%%%%%%%%%%%%

\setcounter{footnote}{0}

\section{Introduction}
Astronomical datasets continue to expand rapidly in size and complexity. Ongoing surveys like the Dark Energy Spectroscopic Instrument \citep[DESI;][]{dey2019overview} already encompass millions of objects and future surveys, like the Vera C. Rubin Legacy Surveys of Space and Time \citep[LSST;][]{Ivezic2009} and Euclid \citep{laureijs2011euclid}, are expected to broaden this scope to include billions of objects. 

A variety of computational approaches have been developed to process the data from these surveys \citep{ivezic2020statistics}. In recent years, a growing subset of these approaches have employed data-driven methodologies from machine learning (ML). To date, these approaches have largely been separated into two different classes: 

\begin{itemize}
    \item \textbf{Supervised} methods leverage labeled subsets of observational data to perform discriminative tasks like galaxy morphology classification, photometric redshift estimation, weak lensing, etc. \citep[for a recent review, see][]{HuertasLanusse}, and have achieved significant progress in data-rich settings. However, these methods are ultimately constrained by the quantity and quality of labelled training samples available and are often exposed to only a small fraction of the potentially available data during training. Additionally, bespoke supervised models need to be retrained/redesigned from scratch for each new task, creating significant computational inefficiencies in the data analysis pipeline. 
    \item \textbf{Unsupervised} methods use clustering, principal component analysis, and other techniques to bypass the need for labelled data. These have been employed for tasks like strong lens detection \citep{cheng2020identifying}, anomaly detection \citep{margalef2020detecting}, etc. However, while they do not rely on labelled subsets of the data, they are still typically task-specific, and they have lagged behind their supervised counterparts in performance \citep{caron2018deep}. 
\end{itemize}

Recently, a new line of inquiry has explored \textbf{self-supervised learning (SSL)} as an alternative. These approaches learn high-quality embeddings  i.e. low-dimensional representations of the objects that preserve their important physical information - in the absence of labeled training data. These embeddings can then be used for a variety of downstream tasks, eliminating the need to retrain bespoke supervised models from scratch for each new dataset or new task. 

This is achieved by training models to perform some surrogate task, such as identifying corrupted pairs or filling in masked subsections of the input data. This in turn produces a high-quality, low-dimensional representation which can be used as a ``foundation'' for downstream tasks; these types of models are therefore often dubbed \textit{foundation models} \citep{bommasani2021opportunities}. In computer vision \citep[CV;][]{he2021masked, Tong2022_vmae} and natural language processing \citep[NLP;][]{radford2019language}, these approaches have already closed the gap with their supervised counterparts; indeed, zero- and few-shot\numberfootnote{In zero-shot learning, the model applies its learned representations to identify or categorize new, unseen data instances, without the need for additional training specifically on these new categories or instances. In few-shot learning, the pretrained model is fine-tuned with a very small dataset related to the new task.} training on the learned representations can even exceed supervised performance, especially in domains in which training large supervised models from scratch is infeasible due to constraints on labelled data \citep{bommasani2021opportunities}. Moreover, recent works have now extended these results into the physical sciences more broadly \citep{nguyen2023climax, mccabe2023multiple, subramanian2024towards}. 

A variety of SSL strategies have already been deployed in observational astronomy. For example, one of the earliest explorations of SSL in the context of astronomical images is the application of the Momentum Constrative pretraining strategy \citep[MoCo v2;][]{he2020momentum} on galaxy images \citep{Hayat2020,Stein2021}. This framework learns embeddings of images by maximizing the similarity of embeddings between different augmented views of the same image while minimizing similarity with embeddings of other images. These embeddings can then be used to predict galaxy redshift \citep{Hayat2021}, perform similarity searches, and search for rare, scientifically interesting objects like strong gravitational lenses \citep{Stein2021b}. Another prominent example in this field is the application of a Bootstrap Your Own Latent \citep[BYOL;][]{grill2020} strategy for galaxy morphology classification \citep{walmsley2022towards} to achieve state-of-the-art performance after fine-tuning in the low data regime. 

SSL has also been employed on galaxy spectra. For example, \cite{portillo2020dimensionality} use a variational auto-encoder (VAE) to reduce the dimensionality of galaxy spectra to a small latent space before using a decoder to generate the rest-frame spectrum; the learned latent space then possesses significant intrinsic, rest-frame information about the galaxy spectra, which can be used for downstream tasks like outlier detection, interpolation, and galaxy class classification. Further work \cite{teimoorinia2022mapping, melchior2023autoencoding} add successive improvements to the existing VAE; their embeddings are then similarly useful for downstream tasks like anomaly detection \citep{liang2023autoencoding, liang2023outlier}.

However, despite this recent progress, all of the current SSL approaches in observational astronomy have been limited to embedding objects from a single modality at a time. In an astrophysical context though, there exist a number of complementary observations of the same underlying physical processes; for example, galaxies are often measured using a variety of realizations, including photometry, multi-band images, and optical spectra. As such, a universal foundation model for observational astronomy should be able to simultaneously embed cross-modal realizations of the same physical process into a shared latent space. Then, the learned representations of any given object can be easily searched across different modalities and used seamlessly for a variety of downstream tasks. 

\begin{figure*}
    \centering
    \includegraphics[width=0.98\textwidth]{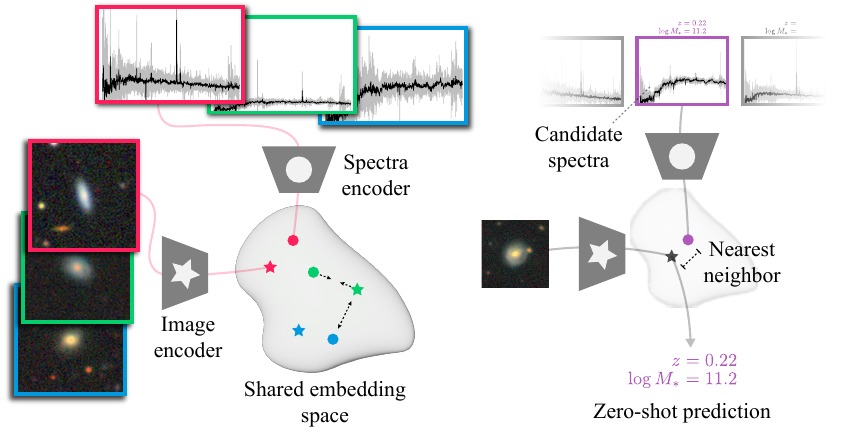}
    \caption{Illustration of the \emph{AstroCLIP} cross-modal training strategy. This approach consists of two steps. First, galaxy images and spectra are embedded separately by pretraining both an image and a spectrum encoder in a SSL setting. Then, these encoders are aligned using a cross-modal contrastive loss. Once aligned, these embeddings allow us to connect and compare cross-modal representations. Moreover, they possess physically meaningful high-level information which can be used for a variety of downstream tasks on which the model was neither trained nor fine-tuned.}
    \label{fig:clip}
\end{figure*}

In this work, we introduce \emph{AstroCLIP}, a cross-modal foundation model for galaxies. Our approach consists of two distinct steps. First, we pre-train state-of-the-art image and spectrum encoders to extract high-quality embeddings of galaxies in a single-modal, self-supervised setting. Then, we align the image and spectrum embeddings by maximizing the similarity between cross-modal embeddings that correspond to the same galaxy while simultaneously minimizing the similarity between cross-modal embeddings that correspond to different galaxies. 

We apply our methodology to optical spectra from the Dark Energy Spectroscopic Instrument (DESI)\numberfootnote{https://data.desi.lbl.gov/doc/} and multi-band images from its corresponding Legacy Imaging Survey\numberfootnote{https://www.legacysurvey.org/}, and demonstrate that our learned embeddings are organized around meaningful physical semantics. This allows them to be used as powerful foundations for both similarity searches and discriminative tasks. This approach is illustrated in \autoref{fig:clip}. Ultimately, we hope that in providing a powerful cross-modal foundation model for galaxy spectra and images, along with a set of physically organized, low-dimensional galaxy embeddings, we will empower a wide variety of downstream data analysis applications in the field.

The main contributions of our work are:
\begin{itemize}
    \item We develop the first self-supervised transformer-based models for galaxy spectra and images. 
    \item We apply a cross-modal training regime to align the pre-trained image and spectrum encoders around shared physical semantics, creating a unified latent space for spectra and images. 
    \item We empirically demonstrate that our cross-modal embeddings capture core physical properties of the underlying galaxies. This enables, with only minimal downstream processing, AstroCLIP to be used for:
        \begin{itemize}
            \item In-modal and cross-modal galaxy similarity searches.
            \item Photometric redshift estimation
            \item Galaxy property estimation from images
            \item Galaxy property estimation from spectra
            \item Galaxy morphology classification from images.
        \end{itemize}
    \end{itemize}
Code for our models, training and testing kit is available online \href{https://github.com/PolymathicAI/AstroCLIP}{here}.

Our paper is organized as follows. In \autoref{sec:ssl}, we provide background on self-supervised learning, as well as on the particular SSL objectives used in the present paper. In \autoref{sec:astroclip_implementation}, we describe the specifics of our AstroCLIP implementation. In \autoref{sec:data}, we provide the data sets that we use to train our models and in \autoref{sec:astroclip_training}, we outline the training process of our models. In \autoref{sec:results}, we present our results on in-modal and cross-modal similarity searches, photometric redshift estimation, galaxy property prediction, and morphology classification. Finally, we discuss our results and further extensions of our paper in \autoref{sec:conclusion}. 

\section{Self-Supervised Learning}
\label{sec:ssl}

In self-supervised learning (SSL), the objective is to train a model to learn to extract rich, low-dimensional representations from data without the need for any labels. This is typically achieved by training the model to perform some contrived surrogate task on the input data. In recent years, a variety of such surrogate tasks have been developed. One common example of such a task in NLP is to predict the next word in a sentence given the previous words; this is typically called autoregressive prediction \citep{radford2019language}. Many other such objectives have been developed, including masked reconstruction \citep{devlin2018bert, he2021masked}, self-distillation \citep{fang2021seed}, and contrastive learning \citep{chen2020simclr, Radford2021}. Ultimately, these approaches have been shown to generate generalizeable, highly informative representations in both NLP \citep[e.g., GPT:][]{radford2019language} and CV \citep{he2021masked, Tong2022_vmae}. 

Despite their task-agnostic training, the zero- and few-shot learning performed on the low-dimensional representations captured by these models has outperformed supervised training in a wide variety of settings, especially in domains in which training large supervised models from scratch is infeasible due to constraints on labelled data \citep{bommasani2021opportunities}. These successes have also highlighted the importance of scale in SSL training strategies, as scaling laws established in both CV \citep{zhai2022scaling} and NLP \citep{fang2021seed} demonstrate significant gains in performance with larger model sizes, dataset sizes, and total compute. 

In the following sections, we outline the relevant SSL training methodologies used in the present paper. In particular, we focus on the contrastive cross-modal strategy that we adopt for AstroCLIP in \autoref{sec:clip}. We then provide a general overview of the self-supervised masked modelling strategy that we adopt for spectrum embedding in \autoref{sec:mim} and the self-supervised self-distillation with no labels strategy that we adopt for image embedding in \autoref{sec:self-dist}; we provide a more detailed description of both these approaches in \autoref{sec:masked-modelling} and \autoref{sec:self-distillation} respectively. For a comprehensive review of self-supervised methods, we direct the reader to \citep{balestriero2023cookbook}.

\subsection{Cross-Modal Contrastive Techniques}
\label{sec:clip}
A variety of techniques have emerged for connecting representations across modalities\numberfootnote{In this context, ``modality'' refers to the type of data input, such as images, textual descriptions, segmentation maps, etc., each requiring different processing techniques.}. One such method, Contrastive Language–Image Pretraining \citep[CLIP;][]{Radford2021}, has achieved widespread success by training neural networks to align language-based descriptions of objects with their corresponding images. CLIP works by using an image embedder and a text embedder to embed both language and image representations into a shared embedding space. These embedders are trained jointly under a contrastive loss, whereby positive pairs (i.e. image-language pairs corresponding to the same object) are brought closer together while negative pairs (i.e. image-language pairs corresponding to different objects) are pushed apart. 

Formally, let $\mathbf{X} \in \mathbb{R}^N$ and $\mathbf{Y} \in \mathbb{R}^M$ be two sets of observations of the same objects from two different modalities; in CLIP, these would be images and textual descriptions corresponding to the same objects. Then, the goal is to construct a pair of encoders, $f_\phi: \mathbb{R}^N \rightarrow \mathbb{R}^d$ and $g_\theta: \mathbb{R}^M \rightarrow \mathbb{R}^d$, that compress these two modalities into a shared $d$-dimensional space. In particular, we want this embedding space to maximize the mutual information between these representations, $I(f_\phi( \mathbf{x}), g_\theta(\mathbf{y}))$. 

Practically, direct computation of the mutual information is a notoriously difficult estimation problem for finite data \citep{mcallester2020formal, song2019understanding}. Therefore, contrastive methods like CLIP typically rely on maximizing approximations of the mutual information. In this case, CLIP uses an Information Noise-Contrastive Estimation \citep[InfoNCE;][]{VanDenOord2018, Gutmann2010}, a variational bound on the mutual information. Although InfoNCE is biased, it represents a stable, low variance bound on the mutual information that has proven successful in a wide variety of contrastive methods \citep{Radford2021}. The InfoNCE loss is given as 
\begin{equation}
\label{eq:infonce}
    \mathcal{L}_{InfoNCE}(\mathbf{X}, \mathbf{Y}) = - \frac{1}{K} \sum_{i=1}^K \log \frac{\exp(S_C(\mathbf{x}_i, \mathbf{y}_i) / \tau)}{\sum_{j}^K \exp(S_C(\mathbf{x}_i, \mathbf{y}_j) / \tau)}.
\end{equation}
Here, $\tau > 0$ represents a smoothing parameter (sometimes referred to as temperature) and $j$ represent the indices of negative examples not associated with the object $i$. 

Additionally, a choice of similarity metric, $S_C$, must be specified to determine the similarity between representations in the embedding space. In CLIP, the cosine similarity between two points in the embedding space is used, such that 
\begin{equation}
\label{eq:sim}
    S_C(\mathbf{x}_i, \mathbf{y}_j) = \frac{(\mathbf{x}_i)^T\mathbf{y}_j}{\|\mathbf{x}_i\|_2^2 \|\mathbf{y}_j\|_2^2}.
\end{equation}
Intuitively, the InfoNCE objective works by bringing together points in the embedding space that correspond to the same underlying physical object and pushing points in the embedding space away from each other if they correspond to different underlying physical objects. Because the InfoNCE loss is itself upper-bounded by the number of negative samples, $\log (K-1)$, CLIP-style models are typically trained with large batch sizes of negative pairs, ranging from $K=512$ to $K=4096$, where larger batch sizes typically correlate with better performance \citep{Radford2021}.

While CLIP has proven successful on a variety of cross-modal problems, the method has shown to suffer from some inefficiencies in training models from scratch, namely due to high computational costs associated with the necessary large batch size and training instability issues when scaling up. Recently however, \cite{sun2023eva} have shown that these issues can be partially overcome using a variety of techniques, including using pre-trained, single modal models as initializers in the CLIP training. 

\subsection{Masked Modelling}
\label{sec:mim}
Masked modelling is an SSL technique used to extract powerful representations in both NLP \citep[Masked Language Modelling, MLM;][]{devlin2018bert} and CV \citep[Masked Image Modelling, MIM;][]{zhou2021ibot} settings. Given an input with random masked patches\numberfootnote{In MLM, the patches of the input are typically contiguous segments of text, while in MIM, the patches of the input are typically square patches of the image.}, the objective in masked modeling is to learn to fill in these randomly masked patches using the remaining unmasked parts of the input. This forces the model to learn to infer the masked patches from the unmasked patches, thereby encouraging robust feature representations of the input that capture the structure and content of the input. Then, when an unmasked input is fed to the model, the learned projection of that input should represent a powerful, low-dimensional embedding. For a more formal discussion, see \autoref{sec:masked-modelling}.

\subsection{Self-Distillation with No Labels}
\label{sec:self-dist}

Self-distillation with No Labels \citep[DINO;][]{caron2021emerging} is another SSL technique widely used in CV and NLP. DINO was inspired by knowledge distillation \citep{buciluǎ2006model}, a method which forces small student networks to approximate the outputs of large, pre-trained teacher networks in order to reduce model size. Like knowledge distillation, DINO still relies on a student network matching the outputs of a teacher network. However, rather than using a pre-trained teacher network, DINO instead uses a copy of the student network composed of an iterated average of past iterations of the student network's weights. By composing the teacher network this way, the teacher network effectively undergoes an ensembling technique, enabling it to guide the student network during training by providing better representation outputs. Since its inception, this technique has been integrated with masked image modeling in \cite{zhou2021ibot}, and further improved with \citep{oquab2023dinov2}, which has demonstrated superior performance on a variety of downstream tasks including semantic segmentation, image classification, video processing, etc. For a more detailed treatment of DINO, iBOT, and DINOv2, see \autoref{sec:self-distillation}.

\section{AstroCLIP Model Implementation}
\label{sec:astroclip_implementation}
The core of our approach lies in the idea that cross-modal observations of a given source can be thought of as filtered, noisy views of the same underlying physical process. Therefore, they should intrinsically possess a shared latent space in which the embeddings of these cross-modal representations can be aligned. To that end, we present a two-step process to train cross-modal galaxy encoders:
\begin{enumerate}
    \item We pre-train two single-modal galaxy encoders separately using SSL techniques. For galaxy images, we pretrain a vision transformer \citep[ViT;][]{dosovitskiy2020image} using a carefully modified version of the DINOv2 self-supervised regime (see \autoref{sec:dinov2}). For galaxy spectra, we pretrain a 1D transformer encoder using a standard mask-filling strategy (see \autoref{sec:mim}). 
    \item We then train (or ``fine-tune'') our pre-trained models in a contrastive setting (see \autoref{sec:clip}) to align the cross-modal embeddings of the same galaxies in a shared embedding space using the CLIP cross-modal alignment strategy (see \autoref{sec:clip}).
\end{enumerate} 
Notably, we opt to pre-train single-modal models separately before CLIP alignment instead of training the entire AstroCLIP model from scratch. For one, the size of the image dataset far exceeds the size of the union between image and spectrum datasets, allowing us to pre-train our image embedder on roughly two orders of magnitude more data. Additionally, previous studies \citep{sun2023eva} demonstrate that the training instabilities and high computational cost associated with CLIP-style training can be partially mitigated by CLIP-aligning pre-trained models. 

We provide the details of the galaxy image and spectrum embedders below. Notably, both models implement transformer-based architectures; we provide extensive background on these in \autoref{sec:transformer_background}, and direct the reader to \cite{vaswani2017attention} and \cite{dosovitskiy2020image} for additional information. We also provide details on the AstroCLIP model implementation. All training details are provided later in \autoref{sec:astroclip_training}.

\subsection{Galaxy Image Transformer}
\label{sec:image_transformer}
Our galaxy image model's architecture is a standard vision transformer \citep[ViT;][]{dosovitskiy2020image}. To prepare a galaxy image, $\mathbf{x} \in \mathbb{R}^{N\times N}$ for the ViT architecture, we first patch the image into non-overlapping, contiguous patches of size $P \times P$. These patches are then flattened, to create a sequence $\mathbf{x}^p \in \mathbb{R}^{K \times (P^2 \cdot C)}$, where $C$ is the number of channels and $K = N^2/P^2$ is the total number of patches, which becomes the effective input sequence length for the transformer. 

Next, we project the patches from dimension $P^2 \cdot C$ to some latent dimension $D_I$ using a trainable, linear projection $\textbf{E} \in \mathbb{R}^{(P^2 \cdot C) \times D_I}$. Additionally, we add position embeddings to each of the patch embeddings; these are standard, learnable 1D vectors $\textbf{E}_{pos} \in \mathbb{R}^{K \times D_I}$ that allow the model to retain positional information for each image patch. Finally, we prepend a class token $\textbf{x}_{\textrm{class}}$ to the sequence. This class token is a learnable embedding that allows the network to aggregate global information in the image, and whose final representation in the network serves as the global image representation. Altogether, this results in a ``processed'' input of 
\begin{equation}
\label{eq:projection_vit}
    \textbf{x}^* = [\textbf{x}_{\textrm{class}}, \textbf{x}^p_1 \textbf{E}, \textbf{x}^p_2 \textbf{E}, ..., \textbf{x}^p_N \textbf{E}] + \textbf{E}_{pos}.
\end{equation}
Once this set of embeddings is generated, we pass them to the transformer model. The transformer consists of a series of Transformer blocks \citep{vaswani2017attention}, each of which apply multi-head cross attention followed by a series of multi-layer perceptron (MLP; sometimes called ``fully-connected'') layers and finally a layer norm. A final layer normalization is applied to the output class and patch tokens. Additionally, we attach a projection head to the class token, which consists of an additional MLP that projects the latent dimensionality of the ViT $D_I$ to some desired dimensionality of the output. We provide the specific implementation details of the galaxy image ViT in \autoref{sec:image_embedder_implementation}. 

\subsection{Galaxy Spectrum Transformer}
\label{sec:spectrum_transformer}
Our galaxy spectrum transformer is loosely modeled after the GPT-2 model, although it performs masked modeling rather than autoregressive prediction\numberfootnote{We deviate from GPT-2 in that we initialize all the weights of the transformer blocks with a normal distribution with standard deviation given by $(2\times \text{fan-in} \times \text{num-layers})^{-1/2}$. The dependence of the standard deviation on the number of transformer blocks is to counteract the effect of having a series of residual connections.} \citep{radford2019language}. As with the galaxy image ViT, to prepare a galaxy spectrum $\mathbf{y} \in \mathbb{R}^{T}$ for the transformer architectures, we first reshape the $T$ dimensional native representation of the spectrum to a sequence of shape $(T \text{ mod } $A$) \times B$, where each element of this new sequence is a contiguous $B$-element segment of the original sequence, and adjacent elements have an overlap of size $A$; these new elements now form our patches, $\mathbf{y} \in \mathbb{R}^{K \times B}$. The patches are once again projected to some latent dimension $D_S$ using a trainable, linear project, and position encodings are added and a class token prepended, as in \autoref{eq:projection_vit}. Once this set of embeddings is generated, we pass them to the transformer model. We provide the specific implementation details of the galaxy spectrum transformer in \autoref{sec:spectrum_embedder_implementation}

\subsection{AstroCLIP Model}
\label{sec:astroclip_model}
The final AstroCLIP model is a sort of compositional model consisting of both the image and spectrum transformers outlined above. The model is constructed using the following steps. First, for any given observation $\mathbf{x}$ with a corresponding label $l = \{\textrm{`image'}, \textrm{`spectrum'}\}$, the model patchifies the input according to the appropriate strategy outlined in \autoref{sec:image_transformer} or \autoref{sec:spectrum_transformer}. Next, the model processes the patchified input through the appropriate image or spectrum transformer, resulting in a processed sequence of vectors with dimensionality equal to the embedding dimension of the transformer, either $D_I$ or $D_S$. 

To transform these vectors into an embedding space that is shared between the image and spectra inputs, AstroCLIP applies a multi-head cross-attention between these final-layer tokens and a learnable query vector $\mathbf{q} \in \mathbb{R}^{512}$. Specifically, the query to this multi-head attention is $\mathbf{q}$, while the keys and values are the final-layer tokens of either the image or vision transformer (see \autoref{eq:attention} for more details). This allows the model to use the attention scores computed between $\mathbf{q}$ and the transformer-output vectors to selectively attend to specific vectors from the transformer output, effectively producing a weighted average of some linear projection of these vectors. The output of this cross-attention is then a single vector $\mathbf{z}^*$ with the same embedding dimension as $\mathbf{q}$; it does not matter how many key and value vectors are received, the dimensionality will remain fixed. $\mathbf{z}^*$ is itself then passed through a series of MLP blocks to produce $\mathbf{z}$.

The final outputs of the AstroCLIP model, $\mathbf{z}$, are embedding vectors of both galaxy images and spectra that reside in a shared, unified latent space. We provide the specific implementation details of the galaxy spectrum transformer in \autoref{sec:clip_implementation}. The alignment of the embedding vector corresponding to a galaxy image, $\mathbf{z}^{\im}$, with the embedding vector corresponding to a galaxy spectrum, $\mathbf{z}^{sp}$, is performed during CLIP training, detailed further in \autoref{sec:astroclip_training}.

\section{Data}
\label{sec:data}
We use galaxy spectra from the Dark Energy Spectroscopic Instrument (DESI) and galaxy images from its corresponding Legacy Imaging Survey (DESI-LS). We use both DESI and DESI-LS for SSL pretraining, along with a variety of additional datasets for downstream tasks. All of these data are detailed below, and a summary of the number of galaxies in each dataset is provided in \autoref{tab:data_points}. 

\begin{table}
\centering
\begin{tabular}{l r}
\hline
\textbf{Dataset} & \textbf{Number of Galaxies} \\
\hline
DESI-LS after Cuts & 76,446,849 \\
Cross-Matched DESI \& DESI-LS  & 197,632 \\
\hspace{3mm} PROVABGS Properties & 105,159 \\
Galaxy Zoo DECaLS Classifications & 222,929 \\
\hline
\end{tabular}
\caption{The number of galaxies present in each of our datasets. In particular, we pre-train our image model on the DESI-LS and our spectrum and AstroCLIP model on the cross-matched DESI \& DESI-LS. We perform downstream redshift estimation on this same dataset, property prediction on the cross-matched PROVABGS dataset, and morphology classification on Galaxy Zoo DECaLS.}
\label{tab:data_points}
\end{table}

\subsection{Self-Supervised Training Datasets}

\subsubsection{DESI-LS Images}
We use the DESI-LS Data Release 9 from January 2021 as prepared by \cite{Stein2021}. The observations in the northern galactic cap (NGC) were captured by the Beijing-Arizona Sky Survey for $g$ and $r$ bands and the Mayall Legacy Survey for $z$ bands respectively, while the observations in the southern galactic cap (SGC) were captured by the Dark Energy Camera Legacy Survey (DECaLS).

We keep every source in the sweep catalogues of the DESI-LS that was not identified as a star and whose magnitude in the $z$-band is between 20 and 21 is kept. After imposing the $\textrm{mag}_z$ cut-off, this results in a total of $76,446,849$ galaxies. Many of these galaxy images include overlapping regions of the sky due to the small angular separation between galaxies in galaxy clusters. 

These galaxies are imaged in three optical bands $(g, r, z)$ at a pixel scale of $0.262$ arcsec. The images extracted by \cite{Stein2021} are taken in $256 \times 256$ cut-outs and we crop these images to $144 \times 144$ center-cuts as the vast majority of galaxies will cover less area than the total size of the cut-outs. Additionally, we normalize the images using a standard Z-scoring regime, whereby we subtract the mean and divide by the standard deviation of the image dataset, ensuring that each pixel value has a mean of 0 and a standard deviation of 1, thus standardizing the input data for consistent model training and performance. 

\subsubsection{DESI Spectra}
We use data from the DESI Early Data Release (EDR) \citep{collaboration2023early}, which consists of spectra observed during the Survey Validation campaign. This campaign was divided into the \textit{Target Selection Validation} phase, designed to finalize target selection, and the \textit{One-Percent Survey}, a pilot survey of the full program that covered roughly $140 \textrm{deg}^2$. Since the dataset includes samples of highly different overall amplitudes, in order to make it easier for the network to process all samples, we Z-score each individual sample. We include the mean ($\mu$) and standard deviation ($\sigma$) information by appending it to the spectrum sequence.

\subsubsection{Dark Energy Survey Image-Spectra Pairs}
We cross-match the DESI-LS galaxy images and DESI spectra using the target IDs associated with each galaxy. This yields a paired galaxy image-spectra sample of 197,632 galaxies. We build this paired sample using the same preprocessing steps for images and spectra detailed above. We split our sample using a 90/10 train-test split for training and evaluation. 

\subsection{Downstream Datasets}

\subsubsection{Photometric Redshift Estimation}
\label{sec:redshifts}
For photometric redshift estimation, we use the catalog-reported redshifts from the DESI spectra associated with each DECaLS image in the cross-matched image-spectrum dataset. We remove spurious entries by only selecting entries for which $\textrm{mag}_g$, $\textrm{mag}_r$, $\textrm{mag}_z > 0$. We split the catalog using the same split as above.

\subsubsection{PROVABGS Catalog}
\label{sec:provabgs}
For galaxy property estimation, we use a sample corresponding to roughly 1\% of the DESI Bright Galaxy Survey. Specifically, we collect estimates of the  stellar mass ($M_*$), star formation rate (SFR), mass-weighted stellar metallicity ($Z_{MW}$), and mass-weighted stellar age ($t_{age,MW}$) from the complementary PRObabilistic Value-Added Bright Galaxy Survey (PROVABGS) Catalog \citep{Chang2023}. In particular, we match our image-spectra pairs with the PROVABGS reported best-fit of the above galaxy properties using the DESI target IDs associated with each galaxy. We remove spurious entries in the PROVABGS catalog by only selecting entries for which $M_* > 0$ and $\textrm{mag}_g$, $\textrm{mag}_r$, $\textrm{mag}_z > 0$. This leaves 105,159 samples, which we split using the same split as above. 

\subsubsection{Galaxy Zoo DECaLS}
\label{sec:galaxy_zoo}
For galaxy morphology classification, we use Galaxy Zoo DECaLS\numberfootnote{https://data.galaxyzoo.org/}. In particular, we use the classifications from GZD-5 \citep{walmsley2022galaxy}, which includes over 7.5 million volunteer response classifications for roughly 314,000 galaxies on a variety of questions, including morphological T-types, strong bars, arm curvature, etc. 

We cross-match the Galaxy Zoo DECaLS galaxies with the DESI-LS. After cross-matching the galaxy databases, we remove any galaxy with fewer than 3 volunteer classifications, resulting in a $1.5\%$ reduction in dataset size. This leaves 222,929 galaxies with associated morphological classifications, which we split using a randomized 80/20 train-test split. 

For each galaxy, we use the debiased\numberfootnote{Debiasing in Galaxy Zoo DECaLS includes both redshift debiasing, which mitigates the debiasing from the fact that higher redshfit galaxies appear fainter and smalleer on the sky, and volunteer weighting, which discards the classifications of volunteers with a reported artificate rate over 0.5 and at least 150 total classifications. For more details, see \cite{walmsley2022galaxy}.} volunteer votes on each of the ten questions. We only use a galaxy to train on a question if 50\% or more of the volunteers shown that galaxy were asked that question. Moreover, we only evaluate on galaxies on which more than 34 volunteers gave classifications, as is convention in \cite{walmsley2022galaxy}. To produce a discrete set of classifications for each of the questions, we round the highest predicted vote fraction for a question to 1, and round the remaining fractions to 0. 

\section{Model Training}
\label{sec:astroclip_training}

As stated above, we train our models using a two-step process; first, we pretrain both image and spectrum transformers in single-modal, self-supervised settings on the DESI-LS galaxy images using the DINO v2 loss and the DESI galaxy spectra using a masked modeling loss respectively. Then, we train the compositional AstroCLIP model on the galaxy image-spectra pairs. 

\subsection{Galaxy Image Pre-Training}

We pretrain the galaxy image transformer on the DESI-LS galaxy images using the DINO v2 self-supervised learning strategy. For each input image, we first create a set $V$ of local and global crops. We use 8 local crops of resolution $60^2$ covering a random square cut-out with area equal to $39.4\%$ of the input image, and 2 global crops of resolution $144^2$ covering a random square cut-out with area equal to $94.7\%$ of the input image. The size of the local crops are chosen such that some part of the target galaxies, which are always centered, is always present in the local crop. The following augmentations are also applied to the various crops:
\begin{itemize}
    \item \textit{Rotation/Orientation}: We randomly flip both global and local crops across both axes with $p=0.5$ probability and randomly rotate the images by a random angle sampled between $\mathcal{U}(0, \pi)$. 
    \item \textit{Gaussian Blur}: We randomly blur each channel of the images using a Gaussian kernel. The Gaussian blur is selected to model additional PSF smoothing, and the size of the blurring kernel is parameterized by lognormal fits to the PSF distribution of the data, as in \cite{Stein2021}. This is applied with $p=1.0$ to our first global crop, $p=0.1$ to our second global crop, and $p=0.5$ to each of our local crops.
    \item \textit{Gaussian Noise}: We randomly add Gaussian noise to the image by sampling the noise level from lognormal distributions tuned for each filter channel, as in \cite{Stein2021}. As with the Gaussian blur, the noise is applied with $p=1.0$ to our first global crop, $p=0.1$ to our second global crop, and $p=0.5$ to each of our local crops.
\end{itemize}
Notably, we opt for far fewer augmentations than the original DINO v2 method - omitting solarization, color jittering, and random grayscale - in order to minimize the total number of physical corruptions applied to our data. 

Once cropped and randomly augmented, we patchify all crops in $V$ using \autoref{eq:projection_vit}. This produces, for each global and local crop, a sequence of patches of length $25$ and $144$ respectively. For the student network, we provide all sequences in $V$, while for the teacher network, we provide only the global crops; thus, the student is fed $25 \times 8 + 144 \times 2 = 488$ patches, while the teacher is fed $144 \times 2 = 288$ patches for each image. The self-distillation loss, $\mathcal{L}_{\textrm{KD}}$, is then computed as the cross-entropy loss between the class token of the student network for its given input and the centered and sharpened class token of the teacher network for its given input; the equation for this loss is provided in \autoref{eq:self-distillation}. Additionally, we apply a random mask to the global crops in $V$ with a masking ratio $r \sim \mathcal{U}(0.1, 0.5)$. We then feed the unmasked global crops to the teacher network and the masked global crops to the student network, and compute the masked-modelling iBOT loss, $\mathcal{L}_{\textrm{MIM}}$, as in \autoref{eq:mim}. Finally, we compute the KoLeo loss for each batch $\mathcal{L}_{\textrm{koleo}}$, as in \autoref{eq:koleo}. 

We train the galaxy image ViT over the entire DECaLS using the composite DINOv2 loss and the procedure outlined above. The exact implementation details of our training are provided in \autoref{sec:image_embedder_training}.

\subsection{Galaxy Spectrum Pre-Training}
We pretrain the galaxy spectrum transformer on the DESI galaxy spectra using the Masked-Modelling self-supervised learning strategy. For each input spectrum, we patchify the spectrum into contiguous, overlapping patches as outlined in \autoref{sec:spectrum_transformer}. We then randomly replace 6 contiguous segments of length 30 (equivalent to length 600 in the original spectra representation) with zeros and train the model to minimize the Mean Square Error loss between the predictions and the ground truth of the replaced segments of the sequence using $\mathcal{L}_{\textrm{MM}}$ provided in \autoref{eq:masked_modelling}. The exact implementation details of our training are provided in \autoref{sec:spectrum_embedder_training}.

\subsection{AstroCLIP Training}
To perform our contrastive training step, we remove the projection head of both the pre-trained image and spectrum transformers and attach the multi-head cross attention described in \autoref{sec:astroclip_model}. We then align both image and spectrum transformers using the InfoNCE loss (see \autoref{eq:infonce}) computed between galaxy images and spectra, where positive pairs are defined as image-spectra pairs corresponding to the same underlying galaxy, and negative pairs are defined as image-spectra pairs corresponding to different underlying galaxies. We use a relatively large batch size of $K = 1024$ image-spectrum pairs to increase the number of negative pairs per batch, as is convention in CLIP-style experiments in computer science \citep{Radford2021}. The exact implementation details of our training are provided in \autoref{sec:clip_training}.

\section{Results}
\label{sec:results}
To demonstrate the capabilities of AstroCLIP, we deploy it across a variety of tasks for which it was neither explicitly trained nor fine-tuned. To that end, we embed the galaxy images and spectra in the various held-out test sets listed above (see \autoref{sec:data}) as follows:
\begin{align}
    \textrm{AstroCLIP} : (\textbf{x}^{\im}, \textbf{x}^{\sp}) \mapsto (\textbf{z}^{\im}, \textbf{z}^{\sp}) \in \mathbb{R}^{512}.
\end{align}
We normalize both image and spectrum embeddings as $\bar{\textbf{z}}^{\im} = \textbf{z}^{\im}/\parallel \textbf{z}^{\im}\parallel_2$ and $\bar{\textbf{z}}^{\sp} =\textbf{z}^{\sp}/\parallel \textbf{z}^{\sp}\parallel_2$. This produces a set of normalized galaxy embeddings in a shared, cross-modal latent space which can easily be queried, searched, and used as summary statistics for the ensuing downstream tasks. 

\subsection{Example Retrieval by Similarity Search}
\label{sec:sim_search}

\newcommand{\heit}{118pt}
\begin{figure*}
    \centering
    % First row of subfigures
    \begin{subfigure}[b]{0.08\textwidth}
        \centering
        \includegraphics[height=\heit]{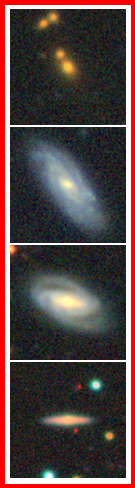}
        \caption{$\mathbf{z}_q$}
        \label{fig:retrieval_1}
    \end{subfigure}%
    \hfill
    \begin{subfigure}[b]{0.22\textwidth}
        \centering
        \includegraphics[height=\heit]{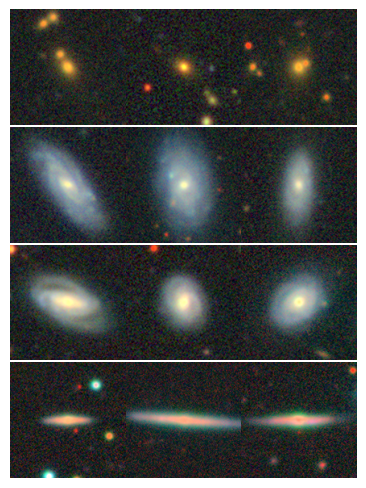}
        \caption{$S_C(\mathbf{z}_q^{\im}, \mathbf{z}^{\im})$}
        \label{fig:retrieval_3}
    \end{subfigure}%
    \hfill
    \begin{subfigure}[b]{0.22\textwidth}
        \centering
        \includegraphics[height=\heit]{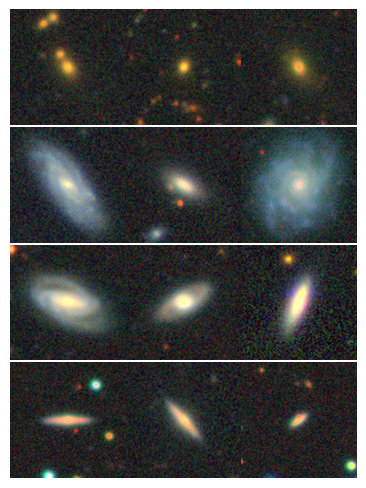}
        \caption{$S_C(\mathbf{z}_q^{\sp}, \mathbf{z}^{\sp})$}
        \label{fig:retrieval_2}
    \end{subfigure}%
    \hfill
    \begin{subfigure}[b]{0.22\textwidth}
        \centering
        \includegraphics[height=\heit]{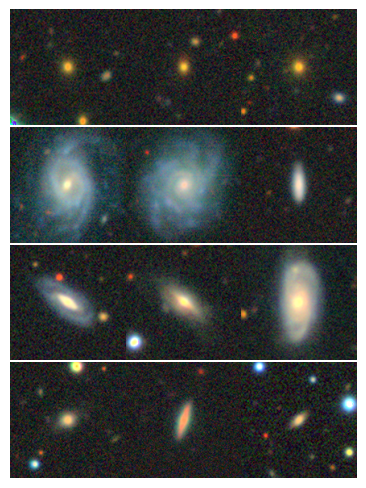}
        \caption{$S_C(\mathbf{z}_q^{\sp}, \mathbf{z}^{\im})$}
        \label{fig:retrieval_4}
    \end{subfigure}%
    \hfill
    \begin{subfigure}[b]{0.22\textwidth}
        \centering
        \includegraphics[height=\heit]{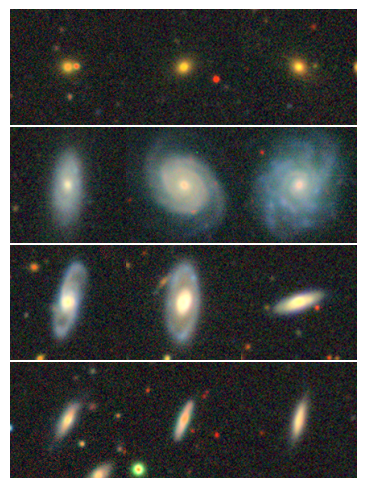}
        \caption{$S_C(\mathbf{z}_q^{\im}, \mathbf{z}^{\sp})$}
        \label{fig:retrieval_5}
    \end{subfigure}
    
    % Add vertical space
    \vspace{7mm} % Adjust the amount of space as needed
    
    % Second row of subfigures
    \begin{subfigure}[b]{0.48\textwidth}
        \centering
        \includegraphics[width=\textwidth]{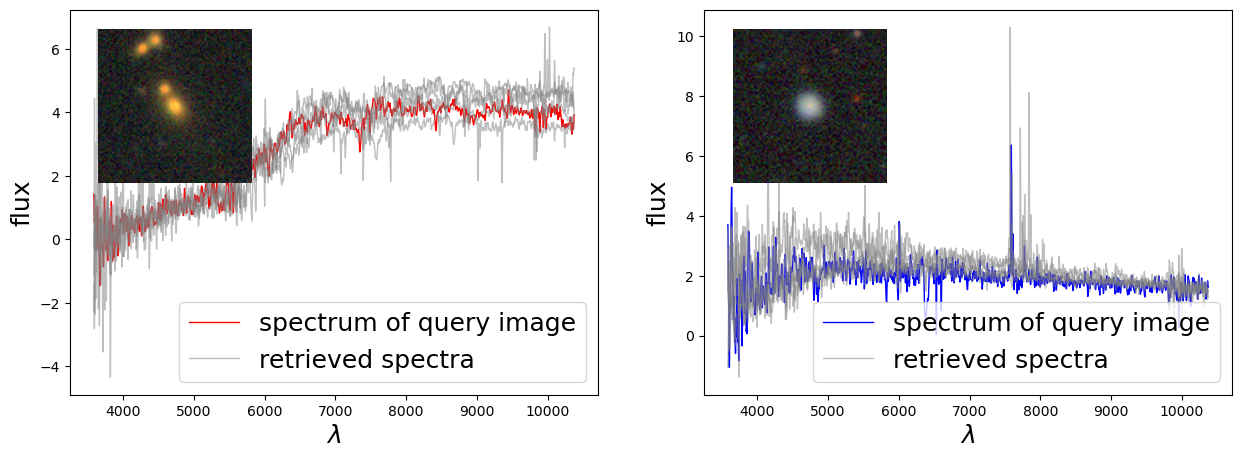}
        \caption{$S_C(\mathbf{z}_q^{\im}, \mathbf{z}^{\im})$}
    \end{subfigure}
    \hfill
    \begin{subfigure}[b]{0.48\textwidth}
        \centering
        \includegraphics[width=\textwidth]{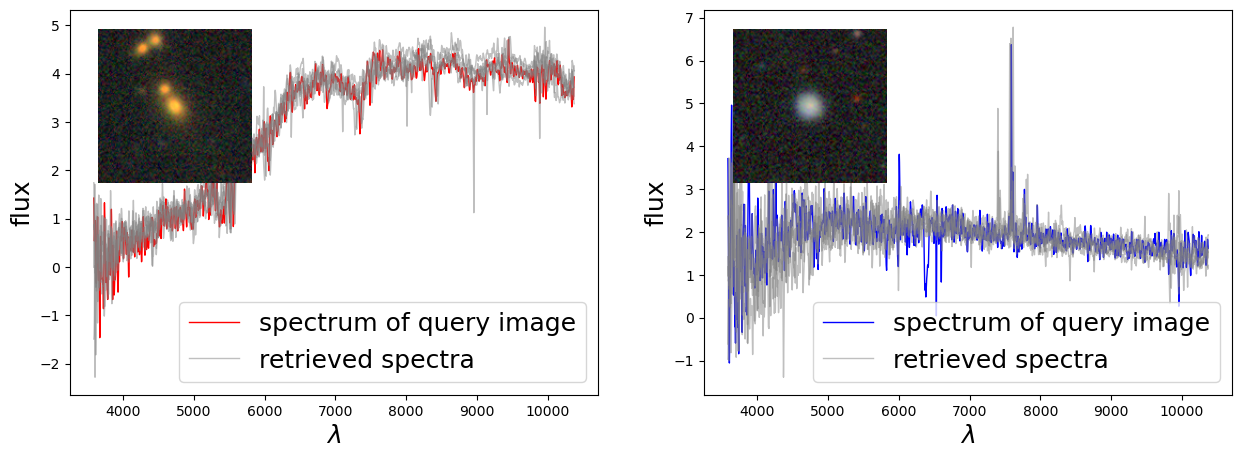}
        \caption{$S_C(\mathbf{z}_q^{\sp}, \mathbf{z}^{\sp})$}
    \end{subfigure}

    \vspace{7mm}

     % Second row of subfigures
    \begin{subfigure}[b]{0.48\textwidth}
        \centering
        \includegraphics[width=\textwidth]{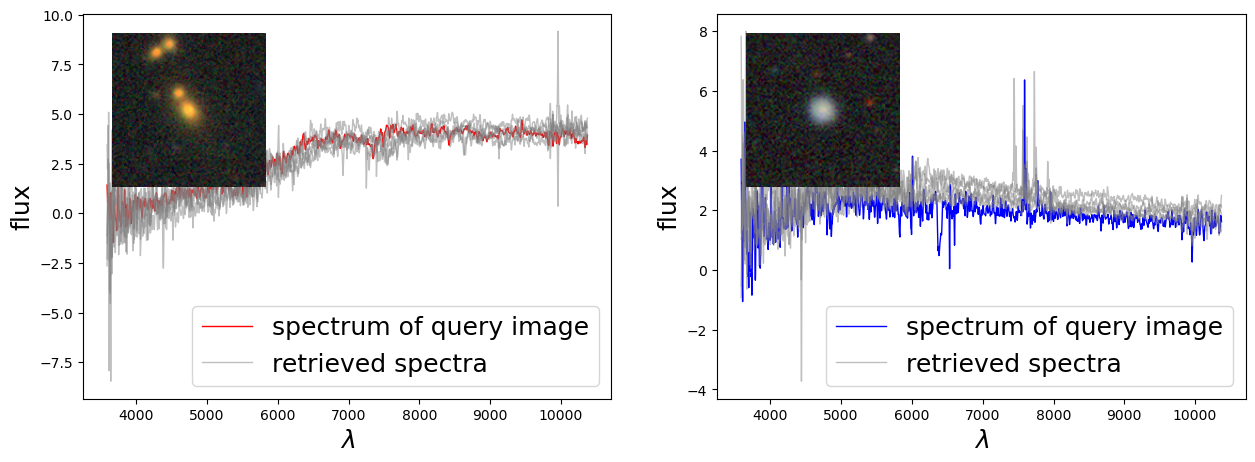}
        \caption{$S_C(\mathbf{z}_q^{\sp}, \mathbf{z}^{\im})$}
    \end{subfigure}
    \hfill
    \begin{subfigure}[b]{0.48\textwidth}
        \centering
        \includegraphics[width=\textwidth]{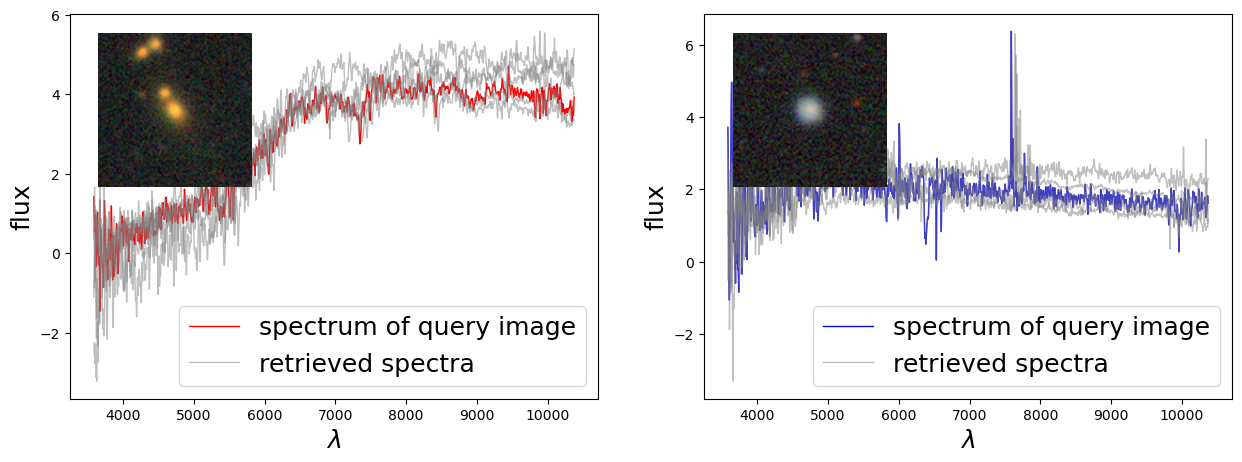}
        \caption{$S_C(\mathbf{z}_q^{\im}, \mathbf{z}^{\sp})$}
    \end{subfigure}
    
    % Global caption for the entire figure
    \caption{Example retrieval from both in-modality and cross-modality searches in the AstroCLIP-aligned embedding space. In particular, for a given query galaxy $\textbf{x}_q$, we embed that galaxy using AstroCLIP as $\mathbf{z}_q = \textrm{AstroCLIP}(\mathbf{x}_q)$ and find the nearest neighbors of that galaxy using the cosine similarity, $S_C(\mathbf{z}_q, \mathbf{z}_{i \neq q})$, between the query embedding and the embeddings of other galaxies in the test set. \textbf{Top:} From left to right, we first show the images of (a) the randomly selected set of query galaxies, and then show the images corresponding to the closest galaxy embeddings using (b) spectrum-spectrum search, (c) image-image search, (d) spectrum-image search, and (e) image-spectrum search. Note that superscripts indicate the input modality. \textbf{Bottom:} We show the retrieved spectra of galaxies nearest to the query galaxy, pictured in each graph, using (f) image-image search, (g) spectrum-spectrum search (h) spectrum-image search, and (i) image-spectrum search. We note that for in-modality searches, the closest neighbor to the query galaxy is by design the query galaxy itself.}
    \label{fig:combined-retrieval}
\end{figure*}

We perform example retrieval using semantic similarity search. Specifically, for some query galaxy, we use its normalized vector embedding to search over all galaxies in the held-out test database. This search is performed using the cosine-similarity (normalized scalar product, see \autoref{eq:sim}) between the embedded query galaxy $\bar{\mathbf{z}}_q$ and all of the other galaxy embeddings in the test database. 

Unlike previous SSL methods in astronomy, AstroCLIP's similarity search is not constrained to a single modality. Instead, because the embedding space produced by AstroCLIP is shared between both images and spectra, both the image and spectrum of any query galaxy can be used to search among all galaxies in the embedded dataset. For example, if we wish to search for galaxy images matching a given query spectrum $\mathbf{x}^{\sp}_{i}$, we simply calculate the cosine similarity between the query spectrum embedding $\bar{\mathbf{z}}^{\sp}_i$ and the image embeddings in the held-out test set, $\bar{\mathbf{z}}^{\im}_j$, and return the target images with the greatest values; no additional transformations or alterations are needed.

We present some examples using this method for both \textbf{in-modality similarity search} - where we determine the neighbors according to the cosine similarity between same-modalitiy embeddings (i.e. $S_C(\mathbf{z}_q^{\sp}, \mathbf{z}^{\sp})$ or $S_C(\mathbf{z}_q^{\im}, \mathbf{z}^{\im})$) - and \textbf{cross-modality similarity search} - where we determine neighbors according to the cosine similarity between cross-modal embeddings (i.e. $S_C(\mathbf{z}_q^{\im}, \mathbf{z}^{\sp})$ or $S_C(\mathbf{z}_q^{\sp}, \mathbf{z}^{\im})$). We present the images of the four ``closest'' galaxies for a randomly selected query galaxy for all four possible pairs of modalities in \autoref{fig:combined-retrieval} (a-e). We also present the spectra of the four ``closest'' galaxies for a red quiescent galaxy and a blue star forming galaxy for all four possible pairs of modalities in \autoref{fig:combined-retrieval} (f-i). By construction, the closest match for an in-modal similarity search is the query itself. Ultimately, this sort of capability is especially important when searching for rare or interesting objects, as exemplified by \cite{Stein2021} paper. 

\begin{figure*}
    \centering
    \begin{minipage}{.6\textwidth}
        \centering
        \begin{subfigure}{\textwidth}
            \includegraphics[width=\textwidth]{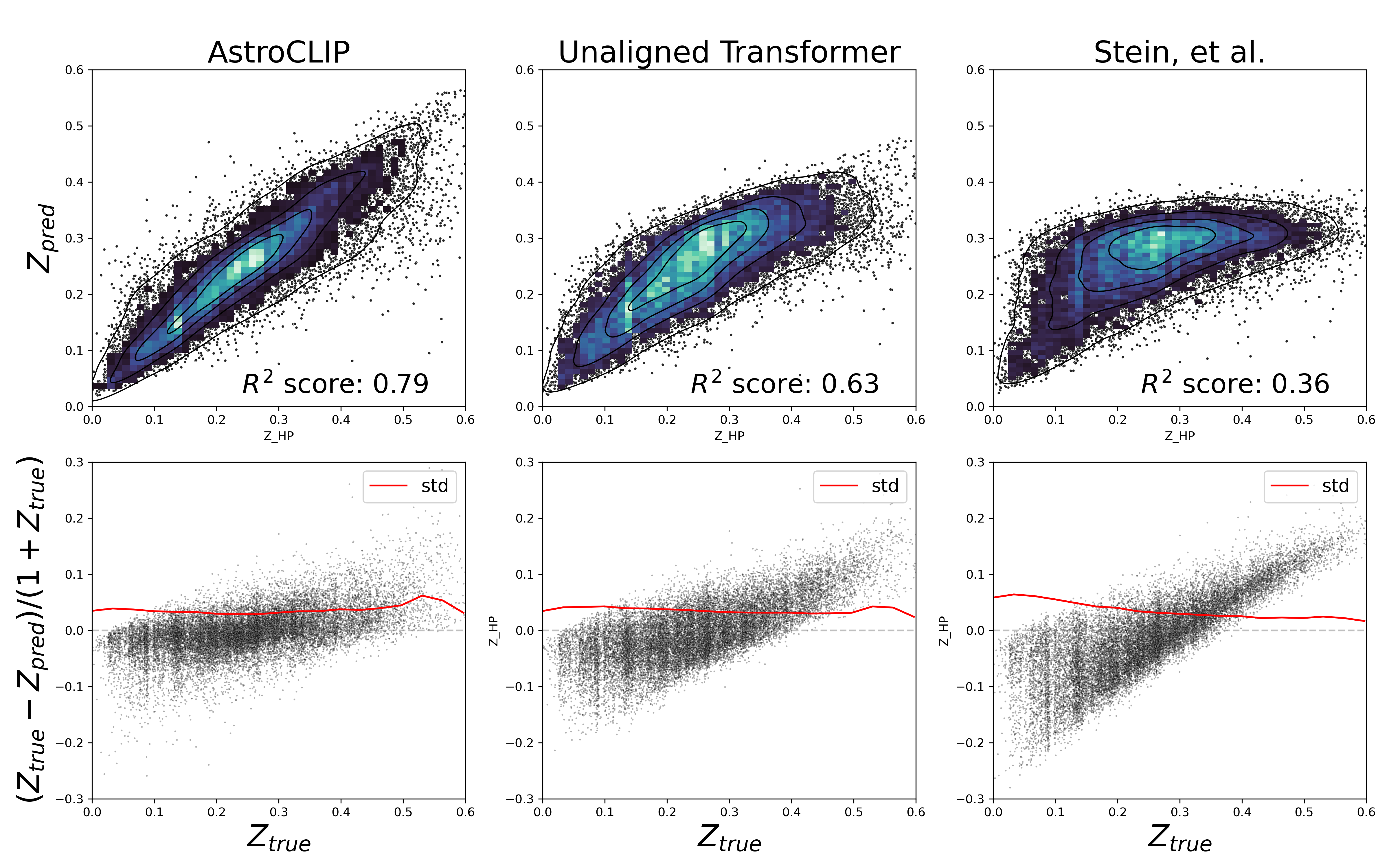}
            \caption{Self-Supervised Zero-Shot Performance.}
            \label{fig:zero_shot}
        \end{subfigure}
        
        \vspace{1em} % Add some vertical space between the figures
        
        \begin{subfigure}{\textwidth}
            \includegraphics[width=\textwidth]{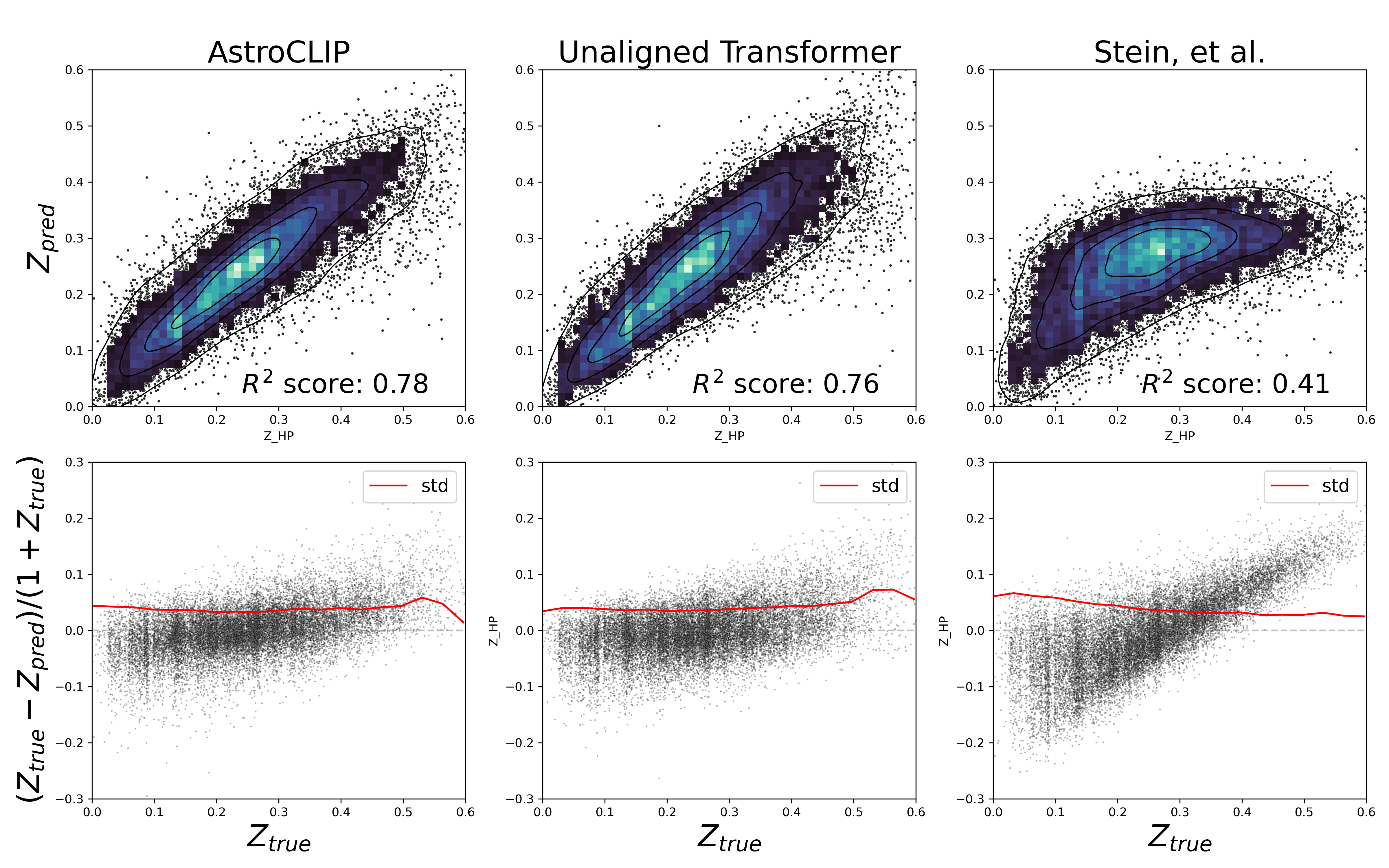}
            \caption{Self-Supervised Few-Shot Performance.}
            \label{fig:few_shot}
        \end{subfigure}
    \end{minipage}
    \begin{minipage}{.38\textwidth}
        \centering
        \begin{subfigure}{\textwidth}
            \includegraphics[width=\textwidth]{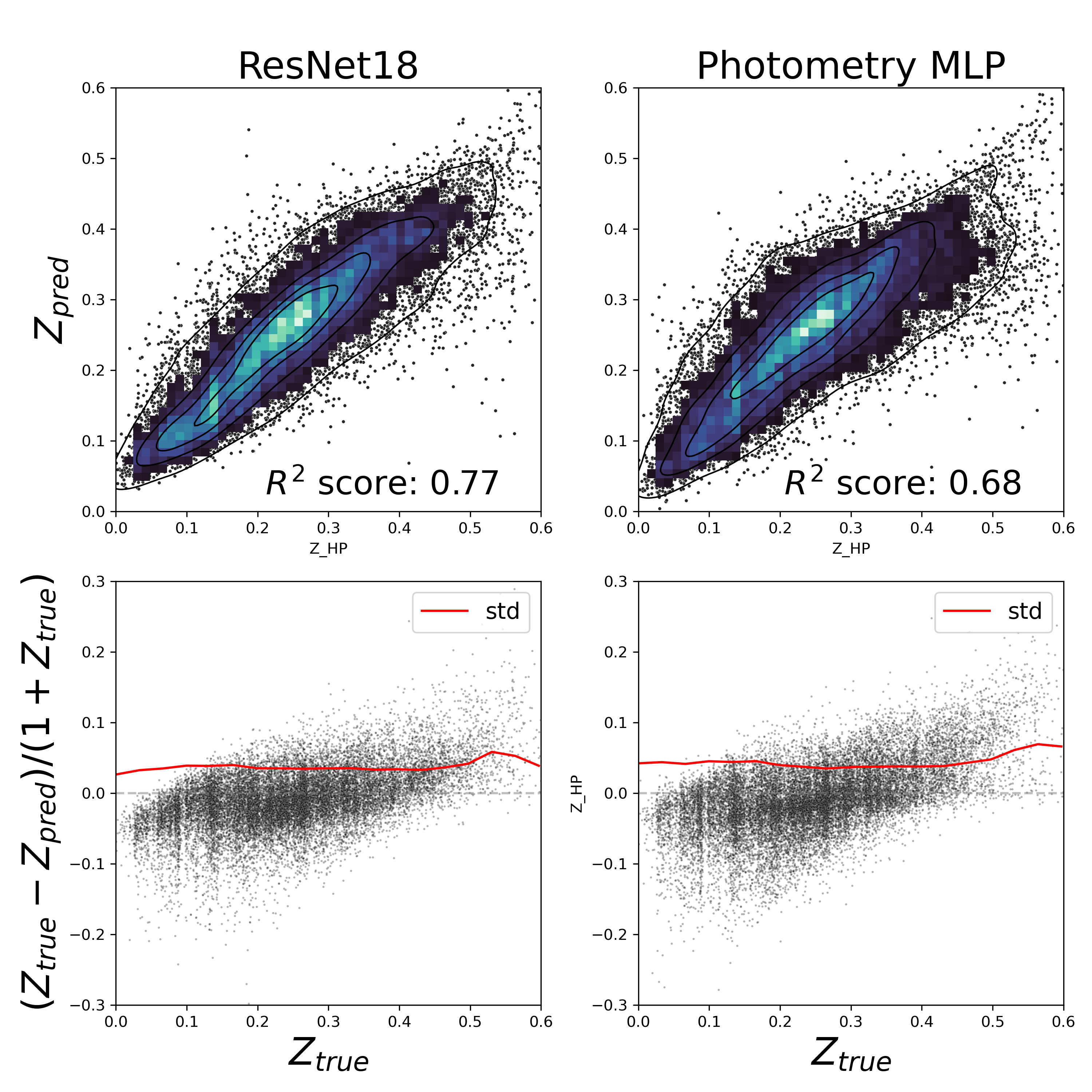}
            \caption{Supervised Baseline Performance.}
            \label{fig:supervised}
        \end{subfigure}
    \end{minipage}
    \caption{Galaxy image redshift prediction and residuals. For zero-shot, we use a simple \(k\)-NN clustering algorithm on the AstroCLIP galaxy image embeddings to predict galaxy redshift. For few-shot, we use a simple MLP to perform the same regression task. We include for comparison the few- and zero-shot performance of our unaligned galaxy image model (DINO) and a state-of-the-art self-supervised model for galaxy images \citep{Stein2021}. We also include two dedicated, supervised, end-to-end models trained on galaxy images (ResNet18) and galaxy photometry (MLP). AstroCLIP performs better than its dedicated, supervised counterpart, despite undergoing no task-specific training or finetuning.}
    \label{fig:redshift_comparison}
\end{figure*}

\subsection{Redshift Estimation}
\label{sec:redshift_estimation}

\subsubsection{Photometric Redshift Estimation}
We evaluate AstroCLIP's performance on photometric redshift estimation. Previous studies have demonstrated that there exists significantly more redshift information in galaxy images than that which would be extractable with simple photometry alone \citep{pasquet2019photometric}. As such, current machine learning methods rely on training dedicated, convolutional neural networks to solve this type of problem, a task which typically involves developing an entire pipeline from scratch and training a dedicated model end-to-end. Because the learned vector embeddings produced by AstroCLIP are already informative about the input galaxies, we are instead able to use simple clustering algorithms (zero-shot) or MLP (few-shot) to extract photometric redshift. Specifically, for zero-shot training, we apply $k$-Nearest Neighbor ($k$-NN) to regress the catalog-reported redshift of a galaxy from AstroCLIP's embedding of that galaxy's image. For few-shot training, we train a single-hidden-layer MLP with width $w=32$ to perform the same regression. We include for comparison the zero- and few-shot results of our unaligned galaxy image transformer model (DINO) as well as those of the single-modal SSL galaxy image model from \cite{Stein2021}. We also include two supervised baselines: a ResNet18 \citep{he2016deep} trained end-to-end on the galaxy images (see \autoref{sec:resnet18}) and an MLP trained end-to-end on the galaxy $(g, r, z)$ photomtetry. 

We report our results in \autoref{fig:redshift_comparison}. In panel (c), we verify that our supervised ResNet18 baseline is indeed able to extract more information than the photometry alone. Overall, AstroCLIP outperforms all models, including the ResNet18 in both zero- and few-shot settings. The strong zero-shot performance of the AstroCLIP model indicates that the galaxy image embeddings are naturally organized in the latent embedding space around galaxy redshift. Contrasting this with the relatively worse zero-shot performance of the unaligned image transformer model, it is clear that the CLIP alignment of the images with the spectra has naturally organized the vector embeddings around galaxy redshift. Given that the spectra are effectively perfectly informative about galaxy redshift, this is to be expected. Either way, both unaligned and CLIP-aligned models outperform the \cite{Stein2021} image embedder, indicating that they are both better organized around (zero-shot) and more informative (few-shot) about galaxy redshift.

\subsubsection{Redshift Estimation From Galaxy Spectra}

\begin{figure}
    \centering    
    \includegraphics[width=0.48\textwidth]{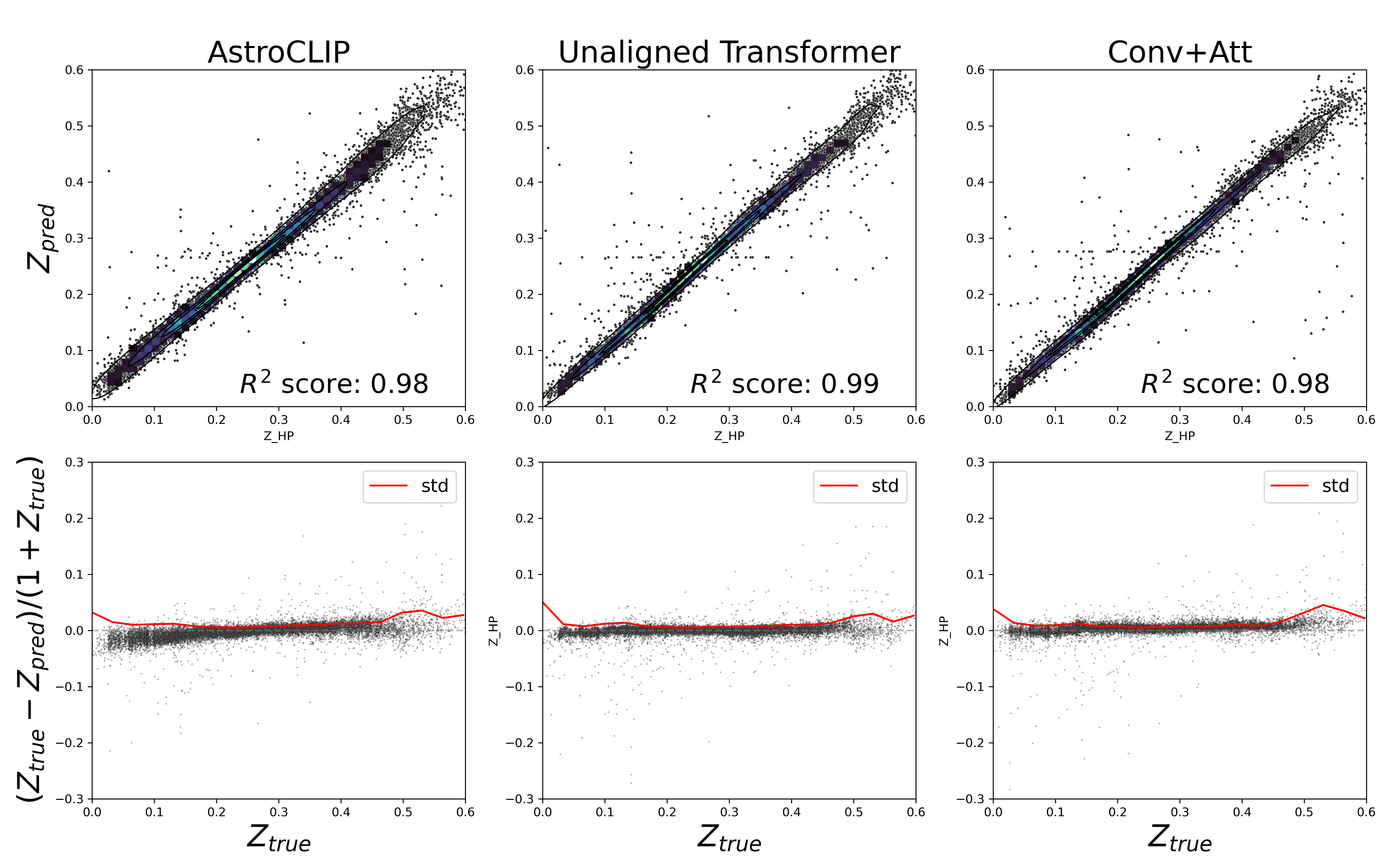}
    \caption{Galaxy spectrum redshift few-shot prediction and residuals. We use a simple MLP trained end-to-end on the AstroCLIP galaxy spectrum embeddings to predict galaxy redshift. Surprisingly, AstroCLIP retains near-perfect redshift information in the spectrum embeddings even after CLIP-alignment with galaxy images. Operating under a pessimistic interpretation of the InfoNCE loss, one would expect CLIP-alignment to only retain the redshift information that is shared by both spectra and images. We also include a supervised Convolutional-Attention Network trained end-to-end for redshift prediction on galaxy spectra}
    \label{fig:redshift_spectrum}
\end{figure}

A galaxy's spectrum should contain near-perfect information on the redshift of that galaxy. This information is accessible with least-square fitting algorithms like \textit{Redrock}\numberfootnote{https://redrock.readthedocs.io/en/latest}, which is used to generate the DESI EDR-reported galaxy redshifts. However, given that galaxy images are not perfectly informative about galaxy redshift, one would expect that the AstroCLIP spectrum embeddings should no longer contain perfect redshift information after CLIP alignment. Afterall, under a pessimistic interpretation of the InfoNCE loss (see \autoref{eq:infonce}), the AstroCLIP model should only be incentivized to keep galaxy redshift information that is shared between galaxy images and spectra. Therefore, there is no reason that it should not discard the redshift information in the galaxy spectra that is not in the galaxy images. Surprisingly, however, this does not seem to be the case. Indeed, in evaluating the few-shot performance of the AstroCLIP spectrum embeddings in \autoref{fig:redshift_spectrum}, we find that there is no material loss of information after CLIP-alignment with the images. This is encouraging, as it means that cross-modal alignment, even with modalities that are not perfectly informative about the the underlying physical process, can still be a good training strategy to generate a model that keeps information. We also compare our results with  a Convolutional-Attention Network based on a state-of-the-art spectrum encoder \citep{melchior2023autoencoding} trained end-to-end on the galaxy spectra (see \autoref{sec:spender}), and find that it is in agreement with the AstroCLIP and unaligned SSL results as well.

\subsection{Galaxy Property Estimation}
\label{sec:physical_properties}
We evaluate AstroCLIP's performance on galaxy property estimation using both galaxy images and spectra as inputs. As above, these tasks are typically performed by dedicated, end-to-end supervised models, whereas here we are able to use simple zero- and few-shot learning on the AstroCLIP embeddings. In particular, we evaluate AstroCLIP's zero- and few-shot performance in estimating the following galaxy properties from the cross-matched PROVABGS \citep{hahn2023desi} catalog detailed in \autoref{sec:provabgs}:
\begin{itemize}
    \item $\mathbf{M_*}$: Stellar Mass
    \item $\mathbf{Z_{MW}}$: Mass-Weighted Stellar Metallicity 
    \item $\mathbf{t_{age}}$: Mass-Weighted Galaxy Age (Gyr)
    \item $\mathbf{sSFR}$: Specific Star-Formation Rate, i.e. star formation activity relative to its stellar mass ($SFR/M_*$)
\end{itemize}
For zero-shot training, we use $k$-NN to regress $[\log M_*, \log Z_{MW}, t_{age}, \log sSFR]$; for few-shot training, we use a single-hidden-layer MLP with width $w=32$ to perform the same regression. 

As in \autoref{sec:redshift_estimation}, we include for comparison the zero- and few-shot results of our unaligned galaxy image (DINOv2) model as well as those of the galaxy image model from \cite{Stein2021}. We also include the zero- and few-shot results of our unaligned galaxy spectrum transformer. Finally, we include three dedicated baselines: a ResNet18 \citep{he2016deep} trained end-to-end on the galaxy images (see \autoref{sec:resnet18}), a Convolutional-Attention Network based on a state-of-the-art spectrum encoder \citep{melchior2023autoencoding} trained end-to-end on the galaxy spectra (see \autoref{sec:spender}), and an MLP trained end-to-end on the galaxy $(g, r, z)$ photometry.

We report our results in \autoref{tab:prediction_power}. Again, AstroCLIP demonstrates an ability to capture in its galaxy embeddings core physical properties of the input galaxy despite undergoing no task-specific training or fine-tuning. For galaxy images, AstroCLIP  outperforms all given baselines, including previous SSL models \citep{Stein2021} and the supervised image (ResNet18) and photometry (MLP) models. For galaxy spectra, AstroCLIP outperforms the supervised photometry baseline on all tasks, and outperforms the supervised spectrum baseline on $M_*$, but performs worse on $Z_{MW}$ and $sSFR$. As above, CLIP-alignment between a less informative (image) and more informative (spectrum) embedding has improved the zero-shot performance of AstroCLIP on galaxy images. However, unlike with redshift estimation, AstroCLIP's performance on spectra has deteriorated relative to its unaligned spectrum transformer model.

\begin{table}
    \centering
    \caption{\textbf{Galaxy property estimation $R^2$ performance.} We present AstroCLIP's zero- and few-shot performance in regressing stellar mass ($M_*$), metallicity ($Z_{MW}$), age ($t_{age}$), and specific-star formation rate ($sSFR$) from galaxy images and spectra. For zero-shot, we use $k$-NN and for few-shot we use a single hidden layer MLP with width $w=32$. We include for comparison the zero- and few-shot performance of our unaligned galaxy image and spectrum SSL transforms (unaligned trans.) and of the SSL galaxy image model from \citep{Stein2021}. We also include three dedicated, supervised baselines trained on galaxy images (ResNet18), galaxy spectra (Conv+Att) and galaxy photometry (MLP). Our models are indicated with an asterisk ($^*$). AstroCLIP outperforms its dedicated, supervised counterparts on most tasks, despite undergoing no task-specific training or finetuning.}
    \begin{tabular}{l l c c c c}
        \toprule
        Source & Method & $M_*$ & $Z_{MW}$ & $t_{age}$ & $sSFR$ \\
        \midrule
        Images & AstroCLIP \\
        & \hspace{3mm} Zero-Shot$^*$ & \textbf{0.74} & \textbf{0.44} & \textbf{0.27} & \textbf{0.44} \\
        & \hspace{3mm} Few-Shot$^*$ & 0.73 & 0.43 & 0.26 & 0.42 \\
        & Unaligned Trans. \\
        & \hspace{3mm} Zero-Shot$^*$ & 0.65 & 0.40 & 0.16 & 0.25 \\
        & \hspace{3mm} Few-Shot$^*$  & 0.72 & 0.43 & 0.23 & 0.40 \\
        & \cite{Stein2021} \\
        & \hspace{3mm} Zero-Shot & 0.43 & 0.30 & 0.11 & 0.20 \\
        & \hspace{3mm} Few-Shot & 0.48 & 0.32 & 0.14 & 0.24 \\
        & ResNet18  & 0.72 & 0.44 & 0.23 & 0.32 \\
        \midrule 
        Spectra & AstroCLIP\\  
        & \hspace{3mm} Zero-Shot$^*$  & 0.87 & 0.57 & 0.43 & 0.63 \\
        & \hspace{3mm} Few-Shot$^*$  & \textbf{0.88} & 0.58 & 0.43 & 0.64 \\
        & Unaligned Trans. \\
        & \hspace{3mm} Zero-Shot$^*$ & 0.84 & 0.57 & 0.38 & 0.62 \\
        & \hspace{3mm} Few-Shot$^*$ & 0.88 & \textbf{0.64} & \textbf{0.47} & \textbf{0.69} \\
        & Conv+Att & 0.85 & 0.62 & 0.43 & 0.67 \\
        \midrule
        Photometry & MLP & 0.67 & 0.41 & 0.27 & 0.34 \\
        \bottomrule
    \end{tabular}
    \label{tab:prediction_power}
\end{table}

\subsection{Neural Posterior Estimation}
We now perform the same set of redshift estimation/property prediction tasks using Neural Posterior Estimation \citep[NPE; \emph{e.g.}][]{rezende2015variational, dinh2016density, papamakarios2016fast, lueckmann2017flexible, greenberg2019automatic}. This enables us to better understand the information content in the AstroCLIP galaxy embeddings. 

Specifically, let $\mathbf{r}$ represent the redshift and property vector for a given galaxy. We are interested in estimating the posterior of $\mathbf{r}$ given the AstroCLIP embedding of that galaxy, $\mathbf{z}$. To that end, we train an ensemble of normalizing flows, $q_\phi$, to estimate the conditional distribution $q_\phi (\mathbf{r} | \textbf{z}) \approx p(\mathbf{r} | \textbf{z})$ over the PROVABGS training set. We provide the relevant background on normalizing flows in \autoref{sec:normalizing_flows} and the details of our implementation in \autoref{sec:nf_implementation}. 

Once trained, if $q_\phi$ represents a good estimate for $p(\mathbf{r} | \mathbf{z})$, we can use it to efficiently calculate the posterior of $\mathbf{r}_i$ given $\mathbf{z}_i$ for some target galaxy $i$. If this distribution is concentrated around the true value, then $\mathbf{z}_i$ is very informative of $\mathbf{r}_i$, while if this distribution is relatively flat around the true value, it is less informative. Repeating this process over the held-out test dataset therefore provides a strong indication of the information content in the AstroCLIP galaxy embeddings. Typically, this is performed by calculating the negative log-likelihood (NLL) over the test set as:
\begin{equation}
    \textrm{NLL} = \frac{1}{N} \sum_{i=1}^K \log q_\phi (\mathbf{r}_i|]\mathbf{z}_i)
\end{equation}

We present the AstroCLIP and baseline NLLs in \autoref{tab:nll_table}. Ultimately, these results corroborate the results presented in \autoref{sec:physical_properties}. Specifically, the AstroCLIP image embeddings once again outperforms both image and photometry supervised baselines, as well as the \cite{Stein2021} SSL model. Interestingly, the AstroCLIP spectrum embeddings also outperform the dedicated spectrum baseline.

In addition to providing a concrete measure of the information content, $q_\phi$ also enables us to efficiently sample from $p(\mathbf{r} | \textbf{z})$. We present sample distributions for a randomly chosen galaxy, along with the true redshift/properties of that galaxy, in \autoref{fig:img_sample_corner} and \autoref{fig:spec_sample_corner} respectively. We also verify that $q_\phi$ is well-calibrated in \autoref{sec:tarp}.

\begin{table}
    \centering
    \caption{Average normalizing flow estimate of the negative log-likelihood (NLL) of the true redshift and galaxy property vector, $\mathbf{r}$, given the input embedding $\mathbf{z}$. In particular, we train an ensemble of normalizing flows $q_{\phi}$ to estimate the conditional distribution $q_\phi (\mathbf{r} | \textbf{z}) \approx p(\mathbf{r} | \textbf{z})$. We then use our normalizing flow to calculate the log-likelihood of the true $\mathbf{r}$ given the input embedding $\mathbf{z}$. AstroCLIP outperforms its dedicated, supervised counterparts on most tasks, despite undergoing no task-specific training or finetuning. Note that lower numbers are better.}
    \begin{tabular}{l l c}
    \toprule
        Source & Method & Neg. Log-Likelihood \\
        \midrule

        Images & AstroCLIP$^*$ & $\mathbf{0.76 \pm 0.00}$ \\
        & Unaligned Transformer$^*$ & $0.81 \pm 0.01$ \\
        & \cite{Stein2021}  & $1.09 \pm 0.04$ \\
        & ResNet18 & $0.77 \pm 0.00$\\

        \midrule 

        Spectra & AstroCLIP$^*$ & $0.14 \pm 0.03$ \\
        & Unaligned Transformer$^*$ & $\mathbf{0.00 \pm 0.04}$  \\
        & Conv+Att & $0.26 \pm 0.00$ \\

        \midrule
        
        Photometry & MLP & $0.92 \pm 0.05$  \\
        
        \bottomrule\\
    \end{tabular}
    \label{tab:nll_table}
\end{table}

\subsection{Galaxy Morphology Classification}
Finally, we evaluate AstroCLIP's performance on galaxy morphology classification. As above, we evaluate the model's few-shot performance by training a 4-layer MLP with width $w=128$ to regress the Galaxy Zoo DECaLS morphology classification from the AstroCLIP image embeddings over the training set. We report the performance of our model on the held-out test set, where we only include performance on galaxies on which more than 34 volunteers provided classifications; this ensures that each answer is well-measured, and is convention in the supervised works from \cite{walmsley2022galaxy}.

We report the accuracy and F1 score for each question over the test set, where we weight the accuracy and F1 score by the support for each class; importantly, not every class has binary classifications, as some classes - like spiral arm count - have up to 6 classes. We include for comparison the few-shot results of our unaligned galaxy image (DINO v2) model, as well as those of the galaxy image model from \cite{Stein2021}. Additionally, we include the reported accuracy/F1-score of the supervised Bayesian classifier from \cite{walmsley2022galaxy}.

\begin{figure*}
    \centering    
    \begin{subfigure}[b]{0.5\textwidth}
        \centering
        \includegraphics[width=\textwidth]{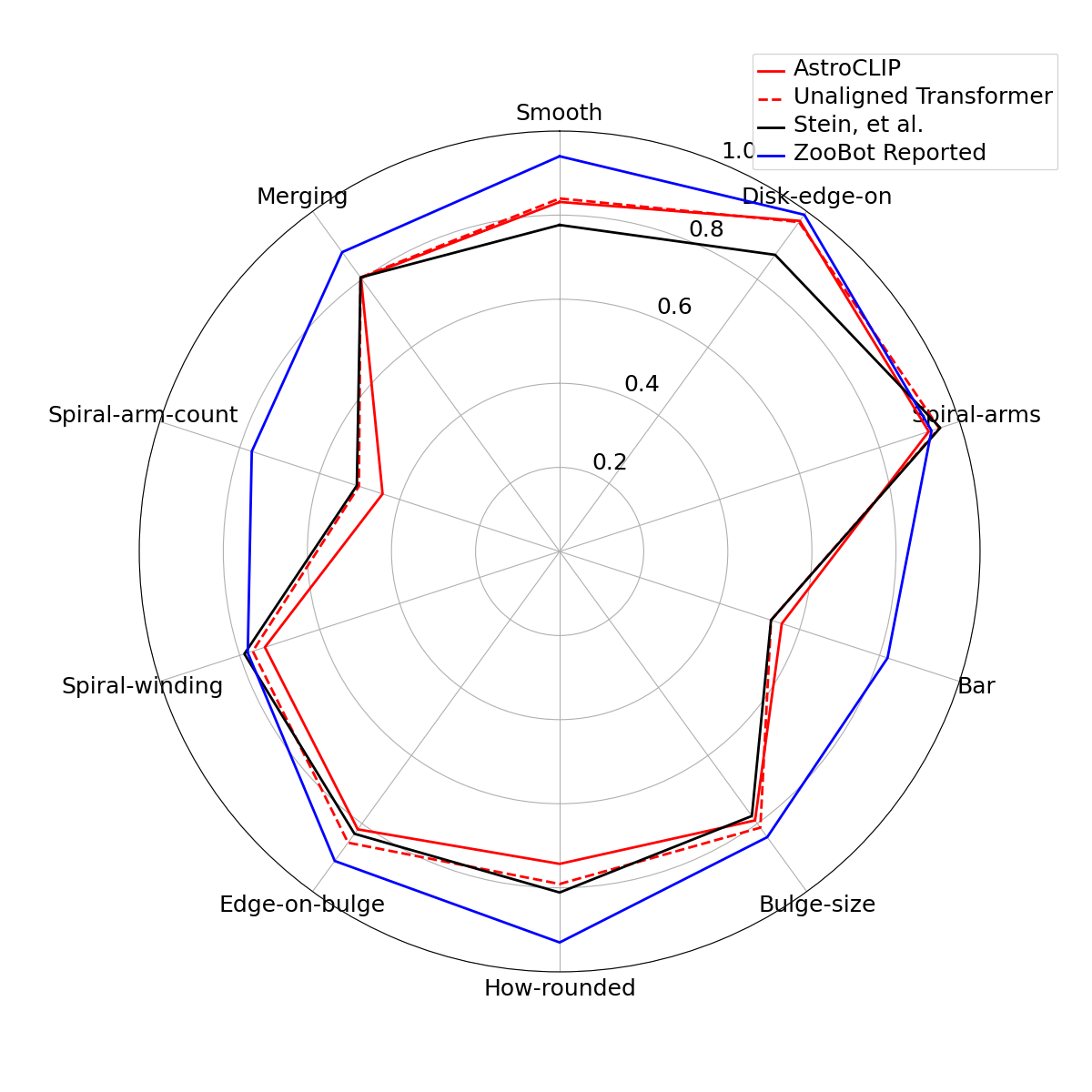}
        \caption{Accuracy.}
        \label{fig:accuracy}
    \end{subfigure}%
    \begin{subfigure}[b]{0.5\textwidth}
        \centering
        \includegraphics[width=\textwidth]{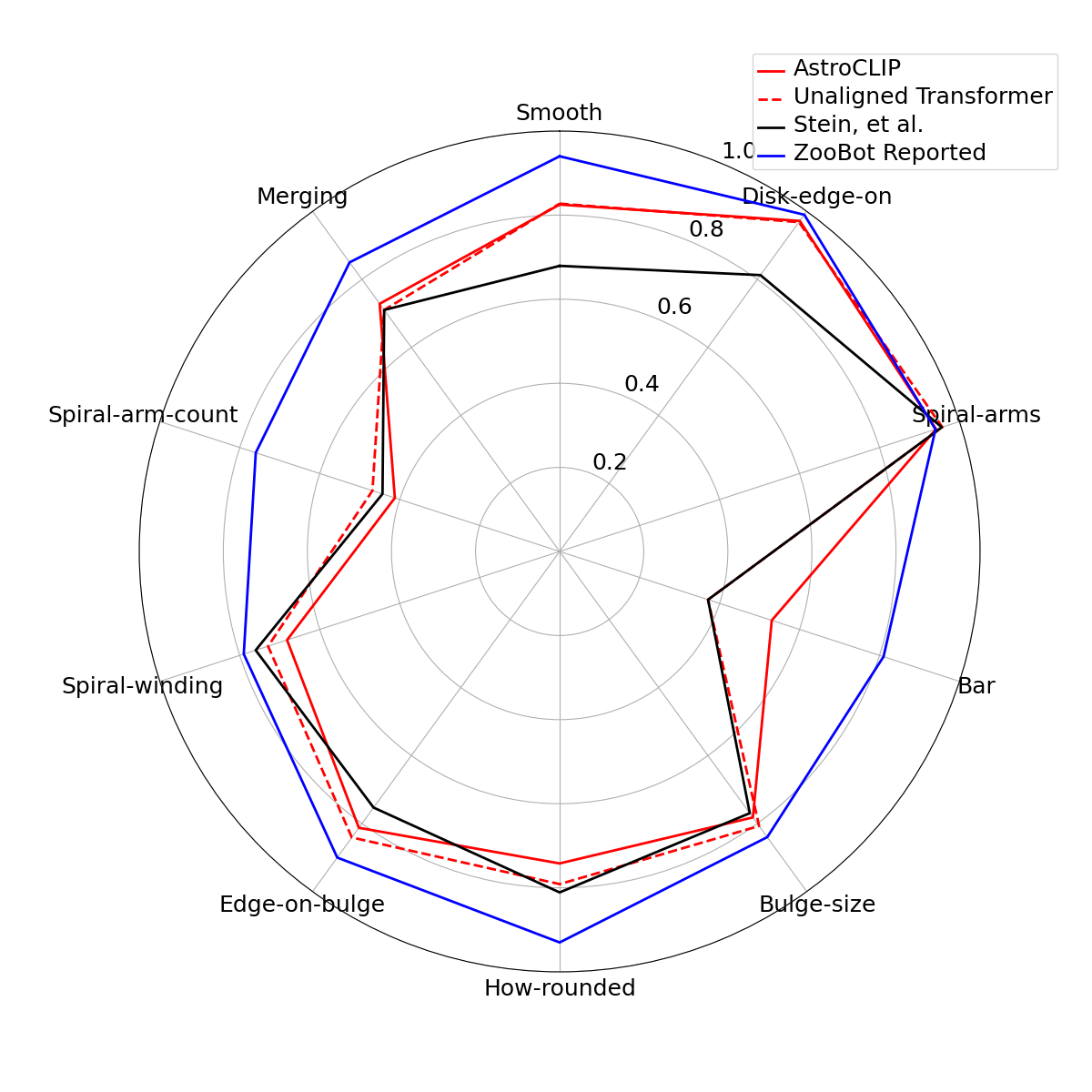}
        \caption{F1 Score.}
        \label{fig:f1_score}
    \end{subfigure}

    \caption{Galaxy morphology classification results. We train a simple MLP on the AstroCLIP galaxy image embeddings to predict the Galaxy Zoo DECaLS GZD-5 morphology classification of that galaxy. We report both the class-weighted accuracy and F1-score of the various models. We also provide the reported class-weighted accuracy/F1-score of the supervised Bayesian classifier \citep{walmsley2022galaxy} for each question. Overall, AstroCLIP achieves relatively strong performance on all questions, and clearly outperforms a state-of-the-art self-supervised model for galaxy images \citep{Stein2021}. Additionally, as in \autoref{fig:redshift_spectrum}, cross-alignment of two different modalities has not materially degraded the performance of the more informative modality, as the difference between the unaligned transformer image model (DINOv2) and AstroCLIP model is negligible.}
    \label{fig:morph_class}
\end{figure*}

We present our results in \autoref{fig:morph_class}. We don't expect any classifier to be able to achieve perfect accuracy on the given tasks, as the volunteer labels themselves possess some intrinsic uncertainty about the underlying galaxy. Therefore, we take the supervised Bayesian classifier as the upper-bound on the achievable accuracy for F1-score of a data-driven model in this particular classification task. We also provide the raw, numerical results for the various models in \autoref{tab:combined_scores}.

Overall, AstroCLIP achieves relatively strong performance on all questions. Raw accuracy score ranges from 97\% on disk-edge-on to 56\% on bar. Overall, AstroCLIP's performance is at least 90\% of that of the supervised model on 6/10 of the questions (disk-edge-on, spiral-arms, bulge-size, edge-on-bulge, spiral-winding, merging). Finally, CLIP-alignment between images and spectra - in this case the less informative modality - has not materially degraded model performance on images; the average accuracy of the unaligned DINOv2 model (78\%) is roughly in-line with that of AstroCLIP (77\%), while AstroCLIP's performance is even slightly better on the disk-edge-on and bar questions. 

\section{Conclusions}
\label{sec:conclusion}
We have presented AstroCLIP, a cross-modal foundation model for galaxies. Our results demonstrate the potential for cross-modal contrastive pre-training to achieve high quality foundation models for astronomical data, which can be used for a variety of downstream tasks. These include accurate in-modal and cross-modal semantic similarity search, photometric redshift estimation, galaxy property prediction from both images and spectra, and galaxy morphology classification. 

Reinforcing our optimism for this approach, our results demonstrate that even if diverse modalities are not all perfectly informative about downstream tasks, the contrastive learning objective is still able to preserve modality-specific information that exceeds that contained in other modalities. This is exemplified by the fact that our spectral embeddings exhibit an emergent ability to retain most of their redshift information while our image embeddings exhibit an emergent ability to retain most of their galaxy morphology information.

Ultimately, we contend that the model's high performance on a wide variety of downstream tasks and its ability to retain modality-specific information are key properties to allow the community to build higher-level models that rely on off-the-shelf astronomical embeddings, just as CLIP language-image embeddings have enabled a wide variety of downstream applications in computer vision.

\section{Acknowledgements}
We gratefully acknowledge the Flatiron Institute for its support. The computations in this work were run at facilities supported by the Scientific Computing Core at the Flatiron Institute, a division of the Simons Foundation. M.P. is supported by the Department of Energy, Office of Science under contract number DE-AC02-05CH11231. This work was granted access to the HPC/AI resources of IDRIS under the allocation 2023-A0151014662 made by GENCI. 

\section{Data Availability}
All the code and resources for this paper will be made available upon acceptance. This includes the code that defines the model, the training codes, and the code used to generate the various results. Additionally, we will publish the trained model which can be used to further evaluate performance or perform additional downstream tasks. The underlying data are all publicly available and all of the software used in this work is open source.



\bibliographystyle{mnras}
\bibliography{bibliography} % if your bibtex file is called example.bib

%\begin{thebibliography}{99}
%\end{thebibliography}

%%%%%%%%%%%%%%%%%%%%%%%%%%%%%%%%%%%%%%%%%%%%%%%%%%

%%%%%%%%%%%%%%%%% APPENDICES %%%%%%%%%%%%%%%%%%%%%

\appendix

\section{Relevant Background}

\subsection{Masked Modelling}
\label{sec:masked-modelling}
As stated above, given an input with random masked patches, the objective in masked modeling is to learn to fill in these randomly masked patches using the remaining unmasked parts of the input. Formally, let us consider an input $\textbf{x}$ composed of a set of $N$ patches, $\{\textbf{x}_i\}_{i=1}^N$. Then, we randomly mask a subset of these patches according to a prediction ratio $r$ to produce $\hat{\textbf{x}} = \{ \hat{\textbf{x}}_i | (1-m_i)\textbf{x}_i + m_i\textbf{e}_{[\textrm{MASK}]}\}_{i=1}^N$, where $\textbf{e}_{[\textrm{MASK}]}$ represents the value of the masked patch and $\textbf{m} \in \{0, 1\}^{N}$ represents the random mask. Let $g_\theta$ be our neural network; then, the projections of each unmasked patch $i$ is $\textbf{x}_i = g_\theta (\textbf{x})_i$ and the projections for each masked patch $i$ is $\hat{\textbf{x}}_i = g_\theta (\hat{\textbf{x}})_i$. The objective in masked modeling is then to minimize the mean-squared error (MSE) loss between the $\textbf{x}_i$ and $\hat{\textbf{x}}_i$ for the same $i$, given as 
\begin{equation}
\label{eq:masked_modelling}
    \mathcal{L}_{\textrm{MM}} = \frac{1}{NK} \sum_{j=1}^K \sum_{i=1}^N \textbf{m}_i \cdot (\textbf{x}_{i} - \hat{\textbf{x}}_{i})^2,
\end{equation}
where $i$ iterates over all of the patches in a given input $\textbf{x}$ and $j$ iterates over all of the $K$ inputs in the training dataset. This forces the model to learn to infer the masked patches from the unmasked patches, thereby encouraging the model to learn robust feature representations of the input that capture the structure and content of the input. Then, when an unmasked input is fed to the model, the learned projection of that input should represent a powerful, low-dimensional embedding. 

\subsection{Self-Distillation with No Labels}
\label{sec:self-distillation}
As stated above, self-distillation with no labels relies on extracting meaningful embeddings by exploiting the dynamics between the training interplay of a ``teacher'' and ``student'' neural network. We first introduce knowledge distillation as relevant background, and then introduce self-distillation as a modification of this framework that enables this type of training in the absence of a fixed, pre-trained teacher network. Finally, we introduce the maskedd image modeling extension proposed by \cite{zhou2021ibot}, and its culmination in a unified framework in DINOv2 \cite{oquab2023dinov2}.

\subsubsection{Knowledge Distillation}
Knowledge distillation \citep{buciluǎ2006model} is a type of training regime that has historically been used to train a small student network to mimic the output of a large, pre-trained teacher network. The ultimate goal of this training scheme is to compress the size of the teacher network. 

Concretely, let $f_{t}$ represent the teacher neural network, and $g_{s}$ represent the student neural network; then, the objective in knowledge distillation is to minimize the cross-entropy between the outputs of both networks for the same input $\textbf{x}$, such that:
\begin{equation}
\label{eq:self-distillation}
    \mathcal{L}_{\textrm{KD}} = - \sum_{j=1}^K P_t(\textbf{x}_j) \log P_s(\textbf{x}_j). 
\end{equation}
Here, $K$ is the size of the training dataset, and $P$ represents a sort of probability distribution of the output of $f$ or $g$, which is attained by using a softmax function to normalize the output of the network:
\begin{align}
    P(\textbf{x})_i = \frac{\exp(f(\textbf{x})_i)}{\sum_{k=1}^K \exp (f(\textbf{x})_k)}.
\end{align}
Knowledge distillation is a powerful compression technique, but it is not applicable to SSL training directly, as it relies on a pre-trained fixed teacher network. 

\subsubsection{Self-Distillation}
To enable knowledge distillation in the absence of a pre-trained fixed teacher network, knowledge DIstillation with NO labels \citep[DINO;][]{caron2021emerging}, or self-distillation, has recently been proposed. Rather than distilling knowledge from a pre-trained teacher, self-distillation works by instead distilling knowledge from past iterations of the student network itself.

Concretely, the student and teacher networks share the same architecture $f$, albeit different weights. The weights of the student network, $\theta_s$, are updated via gradient descent, as is typical of machine learning training. However, the weights of the teacher network, $\theta_t$, are not given access to gradient information. Instead, these are dynamically built from past iterations of the student network's weights. This is done using an exponential moving average (EMA) of the student network's weights \citep{he2020momentum}, such that
\begin{align}
    \theta_t \longleftarrow \lambda \theta_t + (1-\lambda) \theta_s,
\end{align}
where $\lambda$ is a tunable hyperparameter commonly referred to as the smoothing or time constant. 

By composing the teacher network as an iterated average of the student network's past weights, the teacher network effectively undergoes a ensembling technique. This type of model ensembling has been well-explored in the literature \citep{ruppert1988efficient}, and has been shown to lead to better performance and generalization in supervised models. In the context of DINO, it too leads to a teacher network that performs better than its student \citep{caron2018deep}. Therefore, the teacher network, like in vanilla knowledge distillation, is still able to guide the student network during training by providing better representation outputs.   

In practice, DINO adds additional elements to this self-distillation scheme. For one, to promote local-to-global correspondence, a set of $V$ different ``views'' are generated for each input, which in the case of DINO is an image. $V$ consists of both ``global'' views, which consist of large crops of the image, and ``local'' views, which consist of smaller crops of the image. The entire set of $V$ is passed to the student, while only the global views are passed to the teacher. The student and teacher must then still generate the same output representation.

Additionally, to prevent a trivial collapse between the representations learned by the student and teacher networks of the inputs, DINO both centers and sharpens the outputs of the teacher network \citep{caron2021emerging}. 

\subsubsection{image-BERT Pre-Training with Online Tokenizer}
While not originally introduced in a self-distillation context, MIM (see \autoref{sec:mim}) has been extended to this regime in recent works like image-BERT pre-training with Online Tokenizer \citep[iBOT;][]{zhou2021ibot}. Specifically, given some input image $\textbf{x}$, a masked view of the input, $\hat{\textbf{x}}$, is fed to the student network, while the unmasked view $\textbf{x}$ is fed to the teacher. Thus, for any given masked patch $i$, the student network outputs $\hat{\textbf{z}}_s^{i} = P_s^i(\hat{\textbf{x}})$, while the teacher network outputs  $\textbf{z}_t^{i} = P_s^i(\textbf{x}_t)$. These probabilities, like in \autoref{sec:self-distillation}, are once again computed using a softmax function. Then, \autoref{eq:self-distillation} can be easily rewritten as
\begin{equation}
\label{eq:mim}
    \mathcal{L}_{\textrm{MIM}} = - \sum_{j=1}^K \sum_{i=1}^N m_i \cdot P_t^i(\textbf{x}_j) \log P_s^i(\hat{\textbf{x}}_j).
\end{equation}
Functionally, iBOT includes in its loss term some additional complications. For one, iBOT performs MIM on two augmented views of $\textbf{x}$ simultaneously. Then, \autoref{eq:mim} is symmetrized by averaging another cross-entropy term between patches of the other augmented view. Additionally, iBOT includes in its loss another self-distillation-like term between the global representation of the student and teacher network. As in \autoref{sec:self-distillation}, the teachers weights are updated as an EMA of the student weights.

\subsubsection{DINO v2}
\label{sec:dinov2}
self-DIstillation with NO labels version 2 \citep[DINO v2;][]{oquab2023dinov2} is an extension of the DINO self-distillation framework that incorporates the MIM objective from image-BERT Pre-Training with Online Tokenizer \citep{zhou2021ibot} into the DINO objective. For any given input $\textbf{x}$, DINO v2 computes:
\begin{itemize}
    \item The $\mathcal{L}_{\textrm{KD}}$ loss between the features extracted by the student network from both global and local crops of $\textbf{x}$ and the teacher network from the global crops of $\textbf{x}$. 
    \item The $\mathcal{L}_{\textrm{MIM}}$ loss between the randomly masked patches given to the student and the corresponding, unmasked patches given to the teacher. 
\end{itemize}
For both losses, softmax functions are applied to the outputs of the networks, and centering is applied to the teacher outputs to prevent collapse. The composite DINOv2 loss is then given by
\begin{equation}
\label{eq:dino}
    \mathcal{L}_{\textrm{DINOv2}} = w_1 \cdot \mathcal{L}_{\textrm{KD}} + w_2 \cdot \mathcal{L}_{\textrm{MIM}},
\end{equation}
where $w_1$ and $w_2$ are scalars that weight the relative importance of both the DINO and MIM losses. 

In practice, DINOv2 also adds a regularization term to the above composite loss, called the KoLeo regularizer \citep{sablayrolles2018spreading}. This regularizer encourages a uniform span within each batch, and is given by
\begin{equation}
\label{eq:koleo}
    \mathcal{L}_{koleo} = \frac{1}{n} \sum_{i=1}^n \log \min_{j \neq i} \parallel \mathbf{x}_i - \mathbf{x}_j \parallel_2,
\end{equation}
where $n$ represents the total size of the batch. 

Ultimately, the DINOv2 loss has demonstrated superior performance on a variety of downstream tasks including semantic segmentation, image classification, depth estimation, video processing, etc., and a variety of ablation tests have demonstrated the importance of such a composite loss \citep{oquab2023dinov2}.

\subsection{Transformers}
\label{sec:transformer_background}
Transformers \citep{vaswani2017attention} are a type of neural network architecture that employs an attention mechanism \citep{bahdanau2014neural} to attend to and contextualize various parts of its input sequence. 

Here we focus on scaled dot-product attention, a specific implementation of attention, that requires three inputs - queries, $Q$, keys, $K$, and values, $V$. Intuitively, $Q$ represents the set of elements that are seeking information, $K$ represents the elements that are being queried against, and $V$ represents the information that is retrieved based on the similarity between $Q$ and $K$. In this framework, each query in $Q$ is compared against all keys in $K$ to compute a set of attention weights as 
\begin{equation}
    A = \textrm{softmax}\left(\frac{QK^T}{\sqrt{d_K}}\right).
\end{equation}
The output is normalized by the dimensionality of the keys (\(d_K\)) to prevent overly large dot product values, which could lead to gradient vanishing or exploding problems. Ultimately, these attention scores encode how much each value in $V$ should contribute to the output; indeed, the final output of the attention mechanism is computed as a weighted sum of $V$, where the weight of each value is determined by $A$ as
\begin{equation}
\label{eq:attention}
    \textrm{Output} = AV = \textrm{softmax}\left(\frac{QK^T}{\sqrt{d_K}}\right)V.
\end{equation}
When $Q$ and $K$ are the same sequence, the algorithm above is referred to as self-attention, whereas when $Q$ and $K$ are different sequences, the algorithm is called cross-attention. Ultimately, in a simple regression setting, this mechanism allows the network to, for a given entry in the query sequence, ``pay attention'' to the most relevant parts of the rest of the input sequence, calculated in $A$, and use the associated values to produce an output. Moreover, the attention mechanism is permutation-invariant and agnostic to sequence length. 

In practice, trainable weights are applied to the queries, keys, and values before they perform the attention operation. Moreover, multiple attention heads are concatenated at each step, allowing the model to simultaneously attend to information from different representation subspaces at different positions. This is known as multi-head attention, a key feature that enhances the transformer's ability to capture a wide array of dependencies in the input data. Each attention head can be thought of as an independent feature extractor, focusing on different aspects of the input sequence. By concatenating the outputs of these heads, the model integrates diverse perspectives into a comprehensive representation.

Mathematically, the multi-head attention mechanism can be described as follows:
\begin{equation}
\text{MultiHead}(Q, K, V) = \text{Concat}(\text{head}_1, \text{head}_2, ..., \text{head}_h)W^O
\end{equation}
where each head is computed as:
\begin{equation}
\text{head}_i = \text{Attention}(QW^Q_i, KW^K_i, VW^V_i)
\end{equation}
\(W^Q_i\), \(W^K_i\), and \(W^V_i\) are the weight matrices for the queries, keys, and values, respectively, for the \(i\)-th head. \(W^O\) is the output weight matrix that combines the heads' outputs.

In practice, transformers are typically composed of several layers, each containing a multi-head attention mechanism followed by a position-wise fully connected feed-forward network. This design allows the transformer to process all parts of the input sequence in parallel, significantly improving efficiency over architectures that process inputs sequentially. Between each layer, normalization and residual connections are employed to enhance training stability and facilitate the flow of gradients during backpropagation.

\subsection{Normalizing Flows}
\label{sec:normalizing_flows}
Normalizing flows are a type of generative model that are commonly used in Neural Posterior Estimation \citep[NPE; \emph{e.g.}][]{rezende2015variational, dinh2016density, papamakarios2016fast, lueckmann2017flexible, greenberg2019automatic} to estimate either unconditional or conditional probability distributions. These are useful as often times the dimensionality and complexity of the distribution of interest render it impossible to estimate by sampling techniques alone. 

A normalizing flow iteratively transforms a simple multivariate noise source, often the standard multivariate Normal distribution $x \sim \mathcal{N}(\textbf{0}, \textbf{I}_{5 \times 5})$, into the complex parameter distribution $\theta \sim \Theta$ through a series of learned, vector-valued bijective (invertible) transformations $f = f_1 \circ f_2 \circ ... \circ f_n$. This set-up allows them to sample the probability density of the data $\theta$ by simply sampling the latent variable $\mathbf{x}$, and then transforming the variable to $\mathbf{\theta}$ through as $f(\mathbf{x})$. $\Theta$ can then be scored using the multivariate substitution rule as 
\begin{align}
    p(\theta) = \pi(f^{-1}(\theta))\prod_{l=1}^n \left| \det \left(\frac{\partial f_l^{-1} (\theta)}{\partial \theta} \right) \right|, 
\end{align}
where a simple inductive argument is used on $f$. Note that the bijective transformations $f$ must have easy-to-compute Jacobians $\det \frac{\partial f_l (\theta)}{\partial \theta}$ and must be easy to invert for this task to be computationally tractable. 

In many cases, we are interested in the posterior distribution $p(\theta | \mathbf{z})$, where $\mathbf{z}$ is some summary statistic of the data. Luckily, the theory of normalizing flows is easily generalized to conditional distributions, as we simply condition the transformations $f$ on $\mathbf{z}$ to produce the complex, conditionally transformed variable $\theta = f(\mathbf{x})$. Sampling and scoring is analogous to the argument presented above using this conditioning. 

Typically, $f$ is parameterized using a neural network $q_\phi(\theta | \mathbf{z})$, where $\phi$ represents the network parameters. The network parameters are generated by minimizing the Kullback-Leibler (KL) divergence between $p(\theta, \mathbf{z})$ and $q_\phi(\theta, \mathbf{z})p(\mathbf{z})$, which is equivalent to maximizing the log-likelihood over the training set as
\begin{align}
    \phi^* = \textrm{argmax}_\phi \frac{1}{N} \sum^{N} \log q_\phi (\theta_i | \mathbf{z}_i),
\end{align}
where $q_\phi (\theta_i | \mathbf{z}_i)$ is given by the scoring function above. 

\section{Implementation Details}
We provide the implementations details - i.e. the specifics of model architecture, training procedure, hyperparameters, etc. - for the various models trained in the previous sections here. 

\subsection{Galaxy Image Transformer}

\subsubsection{Model Details}
\label{sec:image_embedder_implementation}
While we experimented with various architecture sizes, we find that we achieve best performance when using a ViT-L with patch size $P=12$. Given that our multi-band images have $C=3$ channels, this results in flattened vectors of size $\mathbf{x}^p \in \mathbb{R}^{432}$. We project these to a $D=1024$ dimensional embedding using \autoref{eq:projection_vit}. For our ViT, we use 24 transformer layers, 16 heads in the mult-head self-attention, and MLP hidden layers of width $4 \times 1024 = 4096$. Additionally, we use two separate projection heads for both our student and teacher ViT backbones, each of which has 2048 hidden MLP dimensions and 3 layers. This configuration results in a model with roughly 307 million trainable parameters. 

\subsubsection{Training Details}
\label{sec:image_embedder_training}
Pretraining is performed over 500 epochs on 16 H100 GPUs using the Adam optimizer with a batch size of 96 images per GPU, resulting in a total batch size of 1536. We linearly increase our learning rate from $0$ to $0.004$ over the first 80 epochs of training, after which we decay the learning rate with a cosine schedule. The momentum between student and teacher, $\lambda$, is increased according to a cosine schedule from $0.994$ to $1.0$ during training. We set our weight decay to a fixed value of $0.001$, as we find that increasing our weight decay during training leads to underfitting. We weight the loss between DINO and iBOT losses as one-to-one. Training in this regime takes roughly 48 hours.

\subsection{Galaxy Spectrum Transformer}

\subsubsection{Model Implementation}
\label{sec:spectrum_embedder_implementation}
After experimenting with various patch sizes, we achieve best results when using a patch size of $B=20$ and an overlapping segment of $A=10$. We also experimented with various model architectures, and find that we achieve best performance when using a transformer with with $D=768$ embedding dimensions, 6 transformer blocks, and 6 heads in our multi-head attention module. This results in a transformer with roughly 43.2 million trainable parameters. 

\subsubsection{Training Details}
\label{sec:spectrum_embedder_training}
We pretrain our spectrum encoder on the full DESI spectra dataset using a mask-filling loss. Training is performed on 4 H100 GPUs for a total of 500 epochs, resulting in a total training time of roughly 24 hours. 

\subsection{AstroCLIP Model}

\subsubsection{Model Implementation}
\label{sec:clip_implementation}
In our implementation, we use 4 cross-attention heads followed by two linear layers with 512 embedding dimensions. We also use layer norm and GeLU activation functions. The output of this network is the final embedding of the galaxy, $\mathbf{z} \in \mathbb{R}^{512}$, and should reside in a shared, aligned embedding space after the image and spectrum transformers have been pre-trained and successfully aligned to create AstroCLIP. 

\subsubsection{Training Details}
\label{sec:clip_training}
We train our models on the training split of our paired image-spectrum dataset. We use a queue length of $K = 1024$ image-spectrum pairs. During training, we perform basic data augmentation with random vertical and horizontal flips and random rotations on the images. We train our models using the AdamW optimizer with a base learning rate of 0.0001 with a cosine scheduler and a weight decay of 0.01. We train our model for 500 iterations on a single H100 GPU, which results in roughly 48 hours of training time. Finally, similar to the findings in \cite{girdhar2023imagebind}, we find better performance by fixing the value of the temperature parameter $\tau$ in \autoref{eq:infonce} as opposed to letting it free. We also set the logit scale in our loss to a fixed value of 15.5. 

\subsection{ResNet18 Image Regressor}
\label{sec:resnet18}
We use a modified version of the ResNet18 vision model from \cite{he2016deep}. This model is part of the Residual Network family, known for its ability to train very deep networks through the use of shortcut connections that skip one or more layers. We modify the architecture by changing the first convolutional layer to accept 3-channel $(r, g, b)$ images, and set the kernel size to 7 and the stride length to 2. We also add a final, fully-connected layer that maps the 512-dimensional feature vectors produced by the preceding convolutional and pooling layers to the desired number of output dimensions.

We train two versions of the model: one to regress galaxy redshift, and one to regress the galaxy property vector $\theta = \log M_*, \log Z_{MW}, t_{age}, \log sSFR\}$. We train both over the PROVABGS training set for 100 epochs using the Adam Optimizer and an MSE loss. During training, we prevent model overfitting by applying a number of random augmentations, namely random horizontal and vertical flips with $p=0.5$ and random Gaussian blurring with kernel size 5 and $\sigma \sim \mathcal{U}(0, 2)$. We initialize the learning rate at $\lambda = 5 \times 10^{-4}$. At each epoch, we evaluate the model's performance on the validation set, and take as our best model the model with the best validation performance. We report our results on the held-out test set. We train the model on a single A100 GPU with a batch size of 512, resulting in roughly 1 hour of training time.

\subsection{Spectrum Property Regressor}
\label{sec:spender}
We use a modified version of the time-series encoder from \cite{serra2018towards}. This network first applies four convolutional layers with $[8, 16, 16, 32]$ kernels with PReLU activation functions and dropout. Then, the output of the last convolutional layer split into two halves along its channel dimensions. A dot-product attention is then applied, where one half of the channels serve as the keys $(K)$ and the other half serve as the values $(V)$ for the attention calculation. The attended features are then compressed into a latent representation through an MLP with $[32, 32]$ hidden dimensions. We chose this architecture as it is used as the encoder in a current state-of-the-art spectrum autoencoder setting \citep{melchior2023autoencoding}. 

We train two versions of the model: one to regress galaxy redshift, and one to regress the galaxy property vector $\theta = \log M_*, \log Z_{MW}, t_{age}, \log sSFR\}$. We train both over the PROVABGS training set for 100 epochs using the Adam Optimizer and an MSE loss. At each epoch, we evaluate the model's performance on the validation set, and take as our best model the model with the best validation performance. We report our results on the held-out test set. We train the model on a single A100 GPU with a batch size of 512, resulting in roughly 10 minutes of training time.

\subsection{Normalizing Flow Model}
\label{sec:nf_implementation}
For our problem setting, we use a stack of quadratic rational spline coupling bijections \citep{durkan2019neural} as our bijective transformations $f$. Quadratic spline transformations involve the use of piecewise quadratic functions to create smooth, continuous mappings between variables. We condition these splines on the embedding $\mathbf{z}$ with a fully-connected MLP. For each of our 10 random flows, we randomly choose the number of transformations as $\mathcal{U}\{3, 4, 5, 6\}$ and the random number of MLP hidden dimensions as $\mathcal{U}[32, 128]$. We train each flow using an 80/20 train-validation split on the training set and prevent overfitting by stopping training when the validation log-likelihood has not improved for 20 epochs. We report the negative log-likelihood over the held-out test set as our results. 

\section{Extended Results}
We report a variety of additional results below.

\section{Attention Maps and Performance of Spectrum Encoder}
We look at the attention maps of the cross-attention layer of the spectrum encoder, described in \autoref{sec:spectrum_embedder_implementation}. These plots can help interpret what information the model is looking at when building its representation of the spectrum.

\begin{figure*}
    \centering
    % First row
    \begin{subfigure}[b]{0.5\textwidth}
        \includegraphics[width=\textwidth]{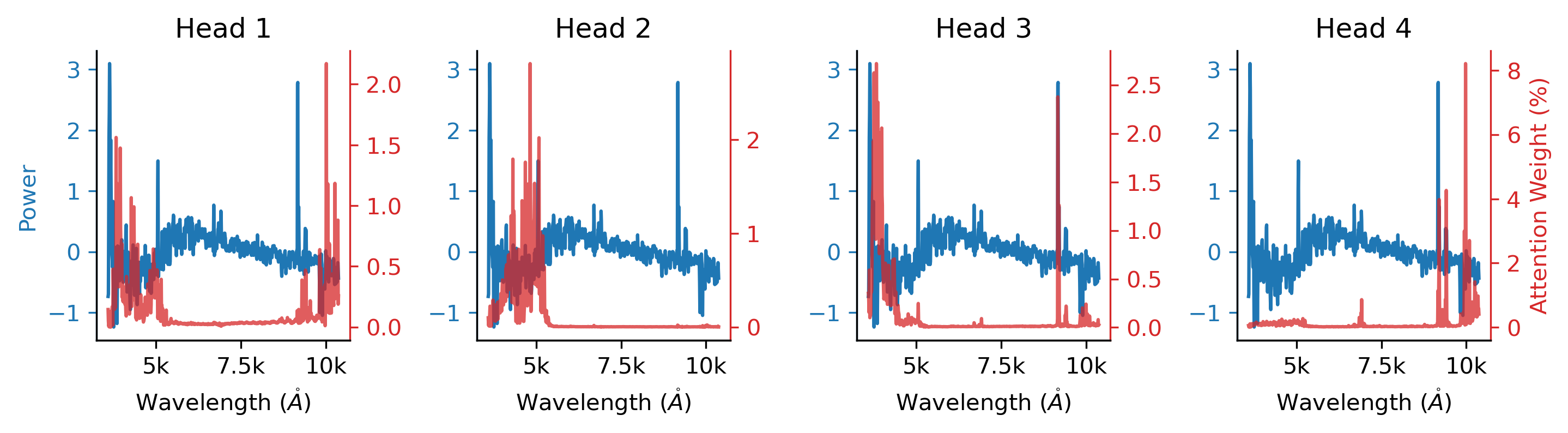}
        \caption{Example 1}
        \label{fig:attention_map_1}
    \end{subfigure}%
    ~
    \begin{subfigure}[b]{0.5\textwidth}
        \includegraphics[width=\textwidth]{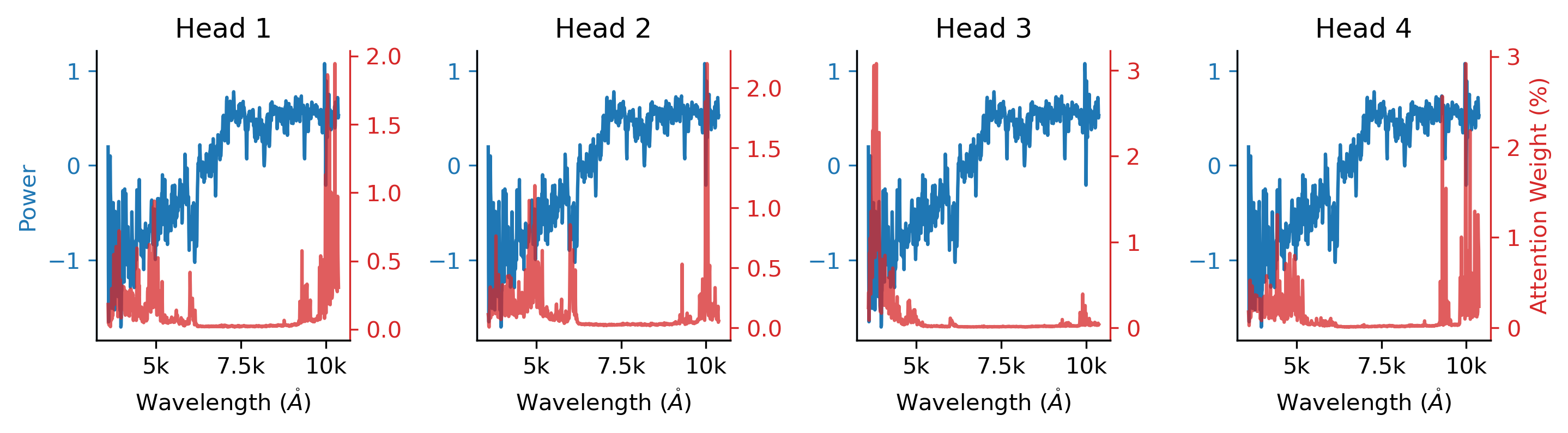}
        \caption{Example 2}
        \label{fig:attention_map_5}
    \end{subfigure}
    
    % Second row
    \begin{subfigure}[b]{0.5\textwidth}
        \includegraphics[width=\textwidth]{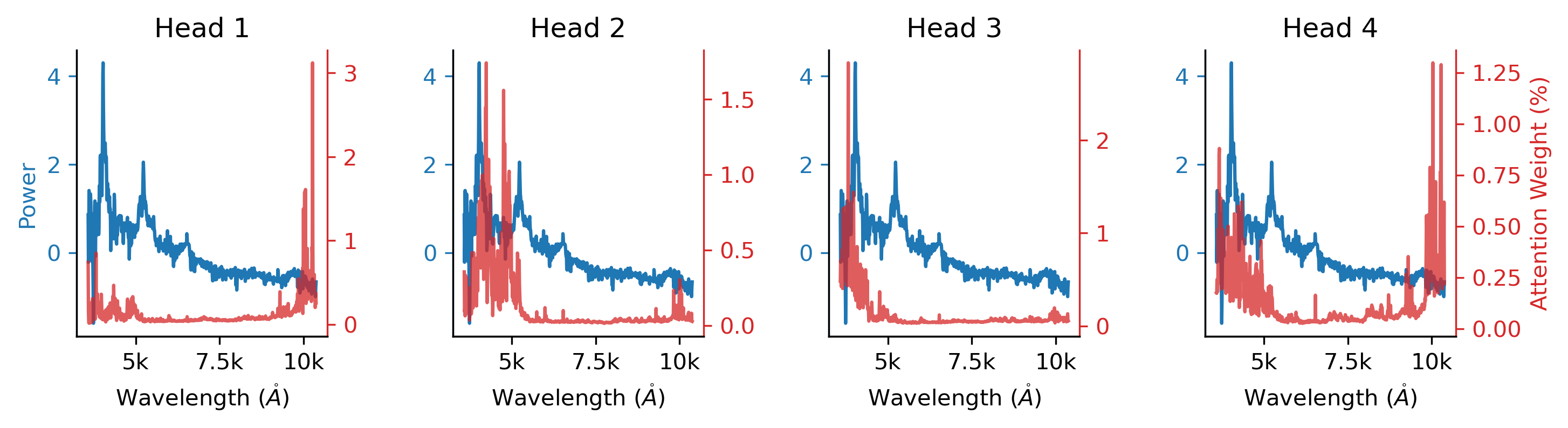}
        \caption{Example 3}
        \label{fig:attention_map_6}
    \end{subfigure}%
    ~
    \begin{subfigure}[b]{0.5\textwidth}
        \includegraphics[width=\textwidth]{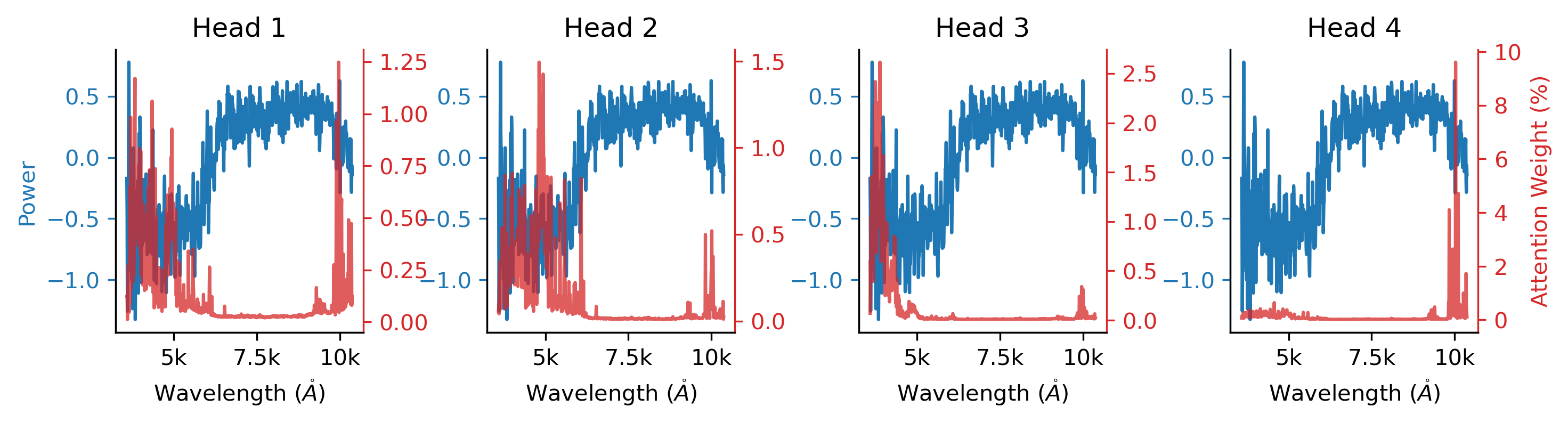}
        \caption{Example 4}
        \label{fig:attention_map_7}
    \end{subfigure}
    
    \caption{Randomly chosen examples of attention maps from the cross-attention layer of the self-supervised spectrum mask-filling model. These visualizations illustrate the model's focus primarily on emission lines within the spectrum, highlighted by pronounced peaks in the attention matrices at these regions, demonstrating the model's ability to identify and emphasize significant spectral features effectively.}
    \label{fig:attention_maps}
\end{figure*}

\begin{figure*}
    \centering
    % First row
    \begin{subfigure}[b]{0.5\textwidth} % Adjust width to fit two columns
        \includegraphics[width=\textwidth]{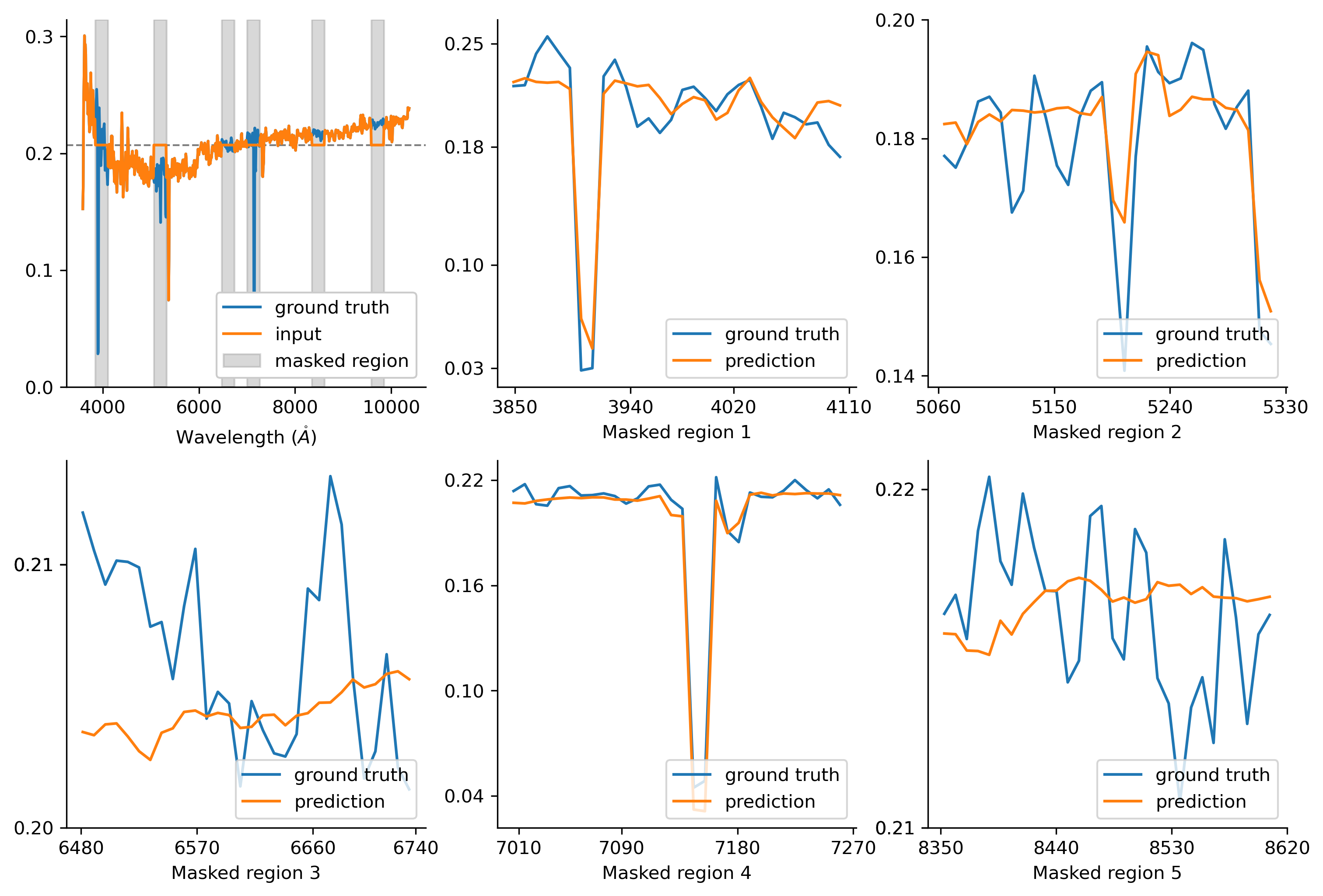}
        \caption{Example 1}
        \label{fig:mask_fill_1}
    \end{subfigure}%
    ~ % This adds a little space between the two figures in the same row
    \begin{subfigure}[b]{0.5\textwidth} % Adjust width to fit two columns
        \includegraphics[width=\textwidth]{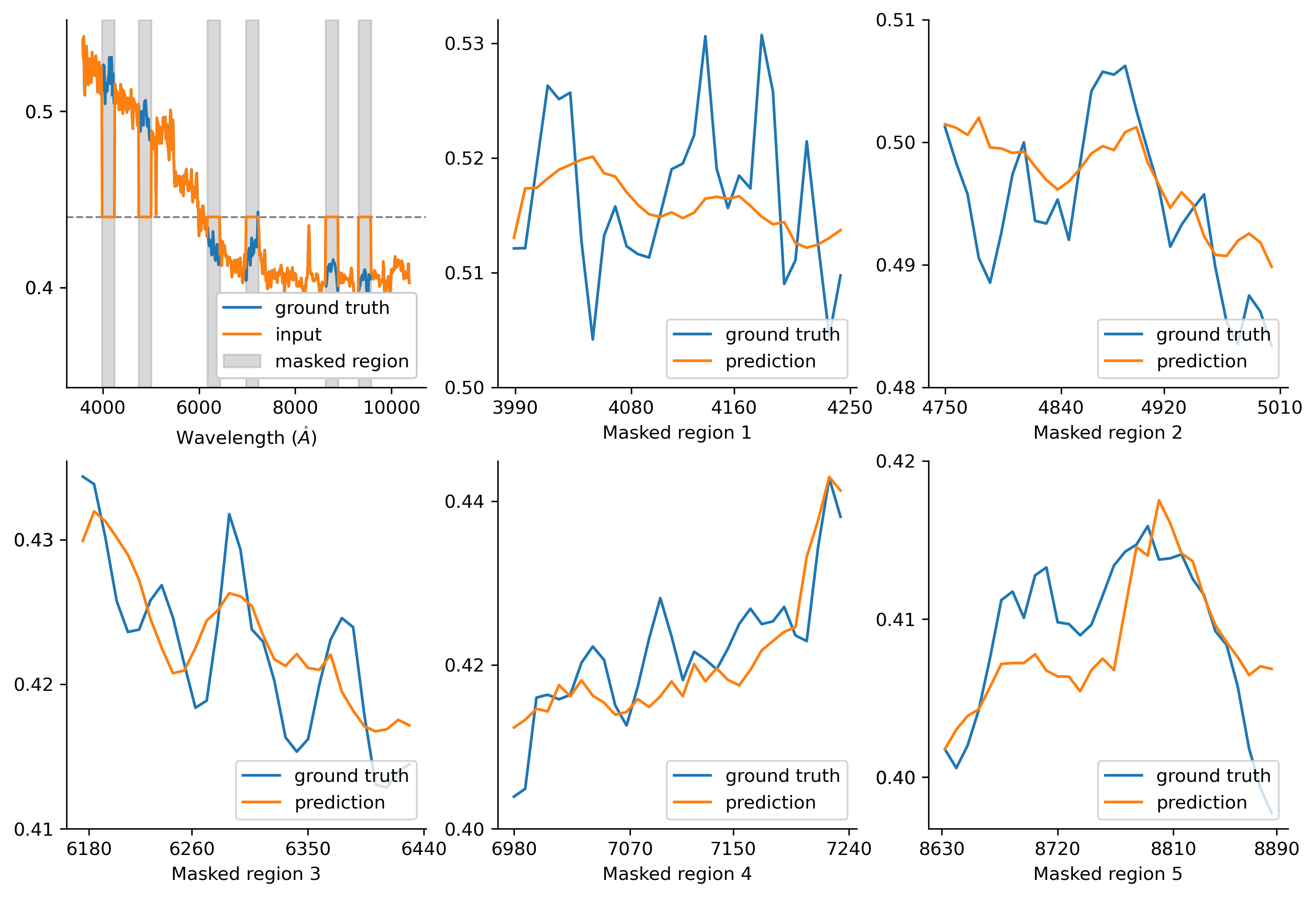}
        \caption{Example 2}
        \label{fig:mask_fill_2}
    \end{subfigure}
    
    % Second row
    \begin{subfigure}[b]{0.5\textwidth} % Adjust width to fit two columns
        \includegraphics[width=\textwidth]{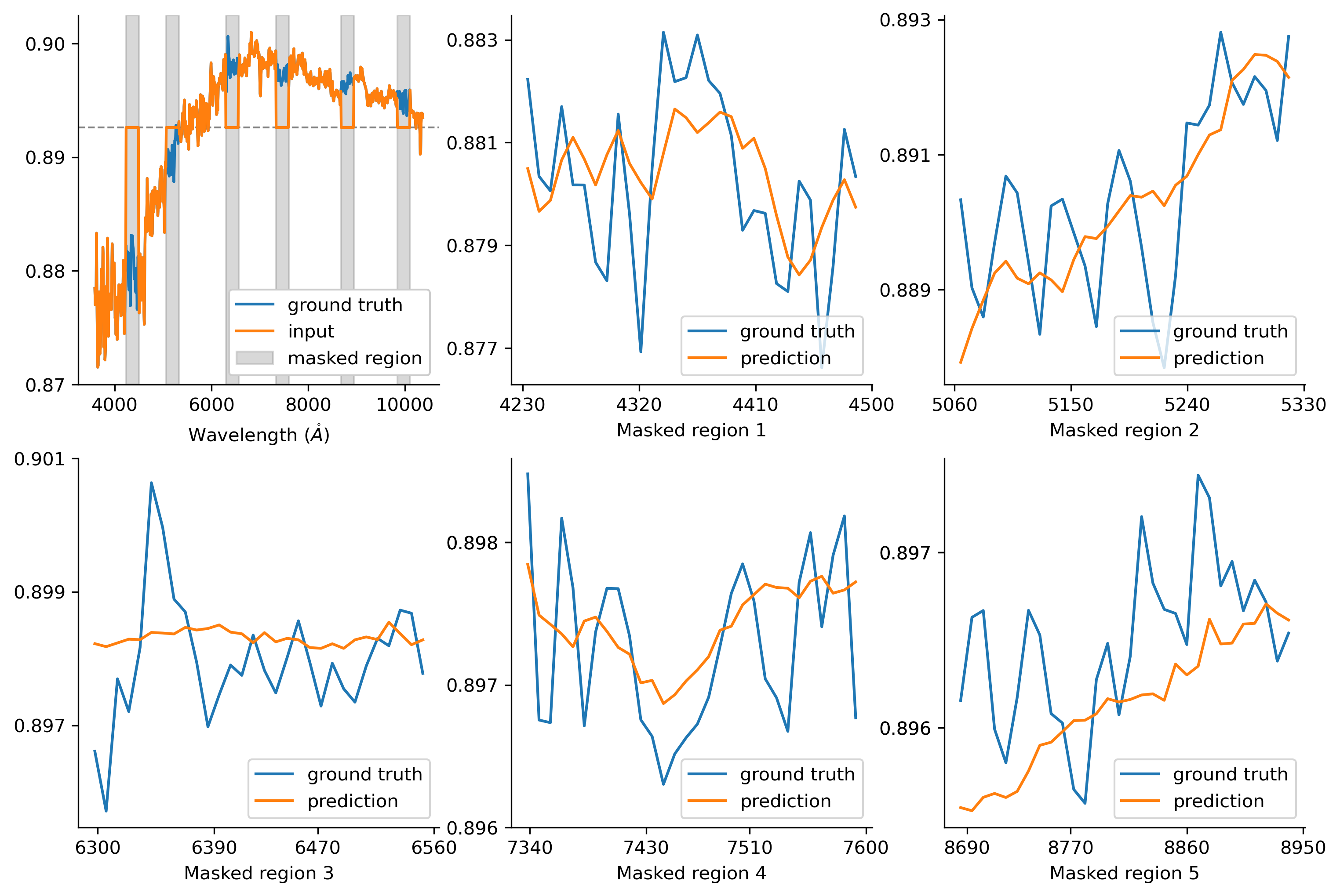}
        \caption{Example 3}
        \label{fig:mask_fill_3}
    \end{subfigure}%
    ~ % This adds a little space between the two figures in the same row
    \begin{subfigure}[b]{0.5\textwidth} % Adjust width to fit two columns
        \includegraphics[width=\textwidth]{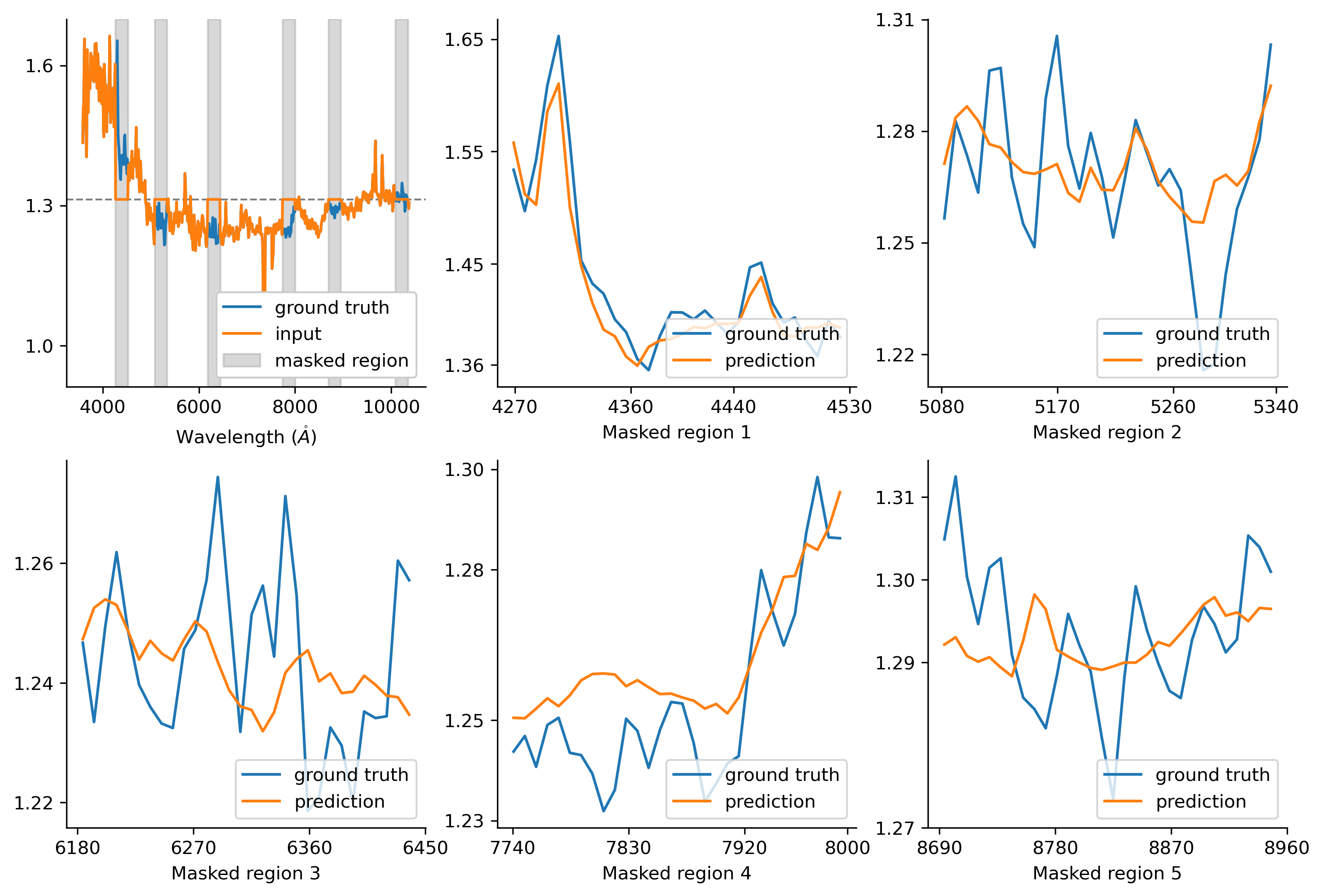}
        \caption{Example 4}
        \label{fig:mask_fill_4}
    \end{subfigure}
    
    \caption{Randomly chosen examples of the performance of the self-supervised trained spectrum mask filling transformer. The spectrum transformer is broadly able to infer the correct shape of missing regions of the spectrum from the broader spectrum context.}
    \label{fig:mask_fill_examples}
\end{figure*}

\autoref{fig:attention_maps} shows a number of examples of these attention maps. We see that the different attention heads have specialized to look for different features. Head 1 seems to be looking at the two extremes of the spectrum which would make it sensitive to different spectral tilts. Head 3 seems to be sensitive to peaks around the 9k$\mathring{A}$ range.  However, it is important to note that this cross-attention layer comes after the 6 layers of self-attention of the pre-trained model. At this stage of the network, information about different sections of the spectrum have likely diffused throughout the entire sequence and therefore the attention maps potentially access information from parts of the spectrum where the attention is zero.

Additionally, we evaluate the performance of the mask-filling model pretrained on the spectra in \autoref{fig:mask_fill_examples}. In these figures, the shaded region denotes the area where the spectrum was zeroed out when passed to the model. The various inserts show close-ups of the smoothed ground-truth (by taking averages of 20 bins) as well as the prediction of the model. We see that the model has learned to reproduce the prominent features of the spectrum. For example, in both \autoref{fig:mask_fill_1} and \autoref{fig:mask_fill_2} a number of the masked regions have fallen on absorption and emmission lines. We see that the model can reproduce these features with high precision.

\subsection{Normalizing Flow Sample Posterior Estimation}
\begin{figure*}
    \centering
    % First subfigure
    \begin{subfigure}[b]{0.5\linewidth}
        \includegraphics[width=\textwidth]{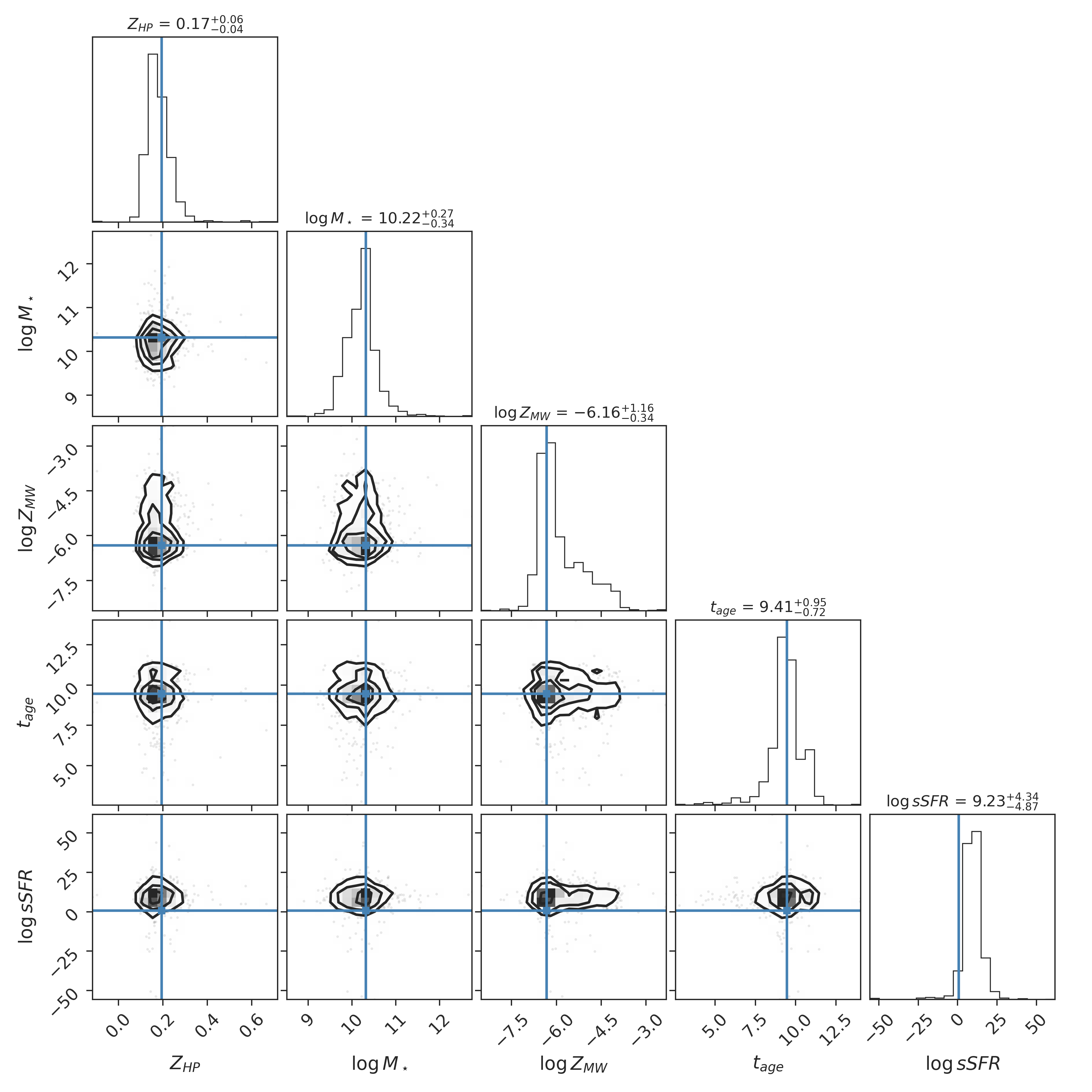}
        \caption{Example Image Input}
        \label{fig:img_sample_corner}
    \end{subfigure}%
    ~ % Space between subfigures
    % Second subfigure
    \begin{subfigure}[b]{0.5\linewidth}
        \includegraphics[width=\textwidth]{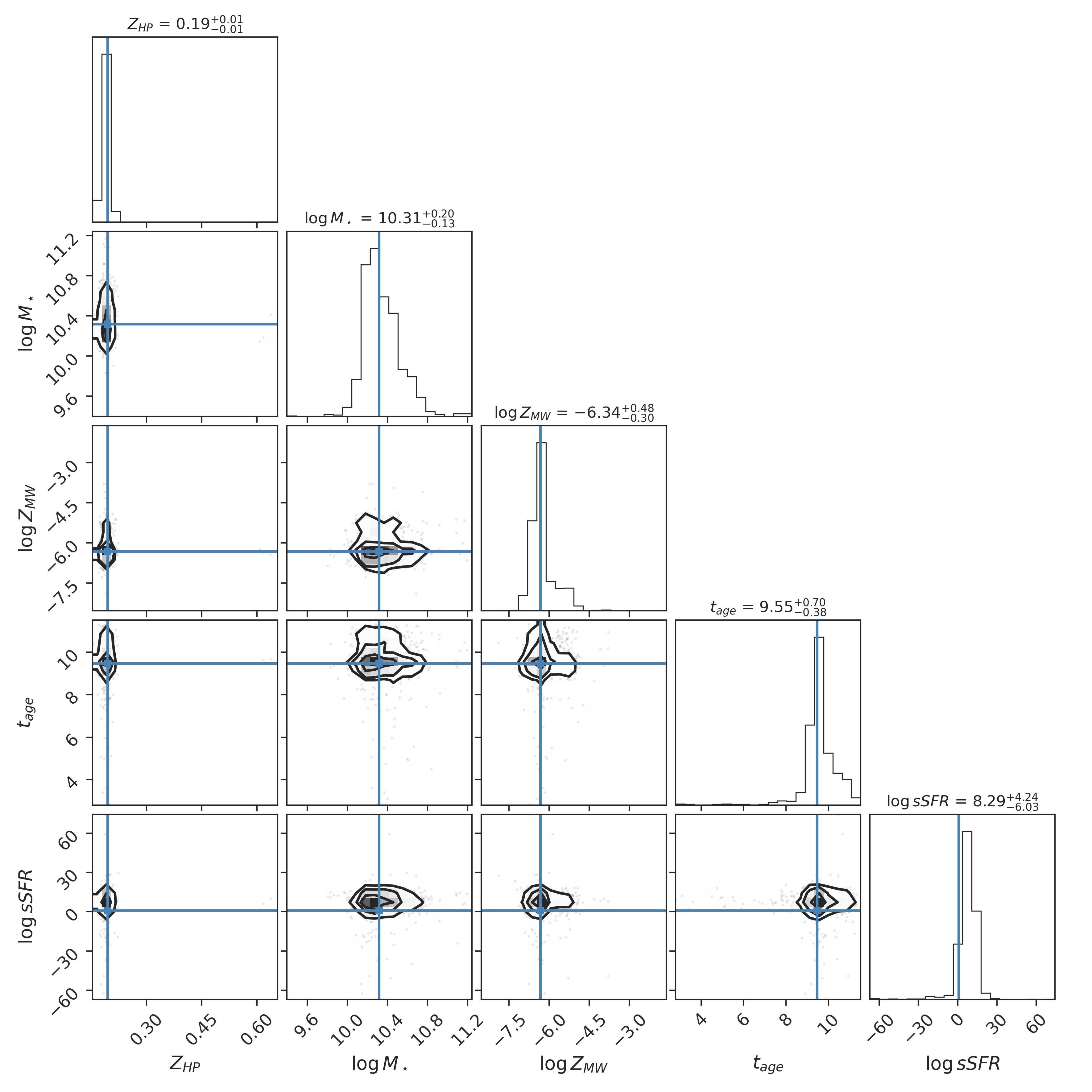}
        \caption{Example Spectrum Input}
        \label{fig:spec_sample_corner}
    \end{subfigure}
    \caption{Galaxy property posterior estimates for a randomly chosen galaxy image and spectrum using normalizing flows. The posterior is estimated using a normalizing flow to map a multivariate Gaussian $\pi = \mathcal{N}(\mathbf{0}, \mathbf{I}_5)$ into the property vector $\mathbf{\theta} \in \mathbb{R}^5$ using learned bijective quadratic splines conditioned on the latent embedding vector $\mathbf{z}^{\sp}$. The flow is then sampled by transforming samples from $\pi$ to $\theta$ using the learned bijective transforms. The true value for each galaxy property is marked with a line in blue.}
    \label{fig:galaxy_properties}
\end{figure*}

We present in \autoref{fig:img_sample_corner} and \autoref{fig:spec_sample_corner} the sampled posterior from our trained normalizing flow $q_\phi (\theta | \mathbf{z})$ for a randomly chosen galaxy image and spectrum respectively, where the flow is conditioned on the AstroCLIP embedding of that image or spectrum. 

\section{TARP Expected Coverage Tests}
\label{sec:tarp}
We ensure that our normalizing flows are well-calibrated using Tests of Accuracy with Random Point (TARP) Expected Coverage Probability (ECP) tests. These have been shown to be necessary and sufficient for exploring the optimality of the posterior estimate \citep{lemos2023sampling}. The TARP method is designed to evaluate the accuracy of generative posterior estimators by creating spherical credible regions centered on a specified random reference point, \(\theta_r\), and then assessing whether these regions capture the true parameter values.

\begin{figure*}
    \centering
    
    \begin{subfigure}[b]{0.8\linewidth}
        \includegraphics[width=\linewidth]{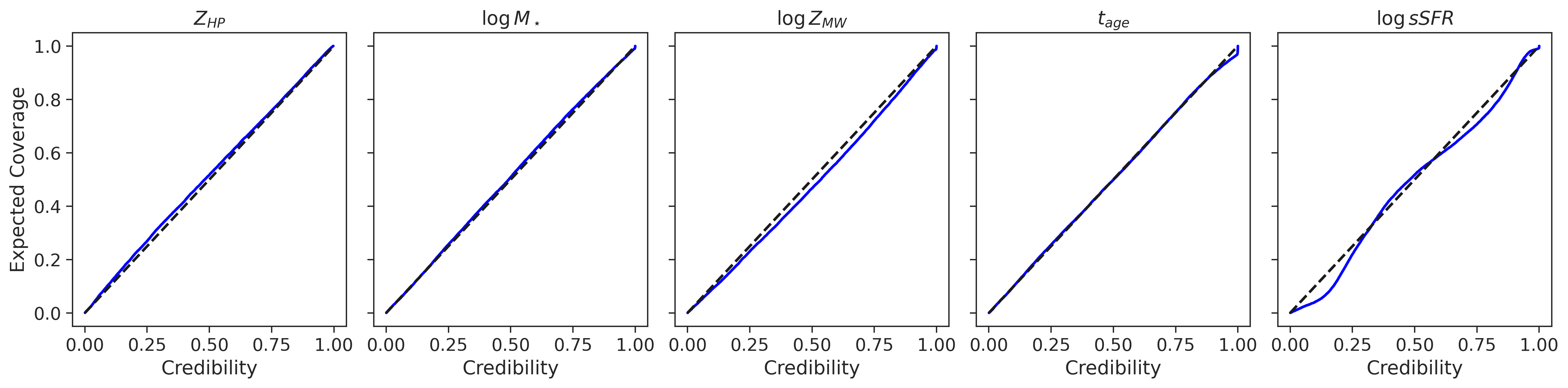}
        \caption{AstroCLIP Image Embedding}
    \end{subfigure}

    \vspace{0.25cm} % adds vertical space between the figures
    
    \begin{subfigure}[b]{0.8\linewidth}
        \includegraphics[width=\linewidth]{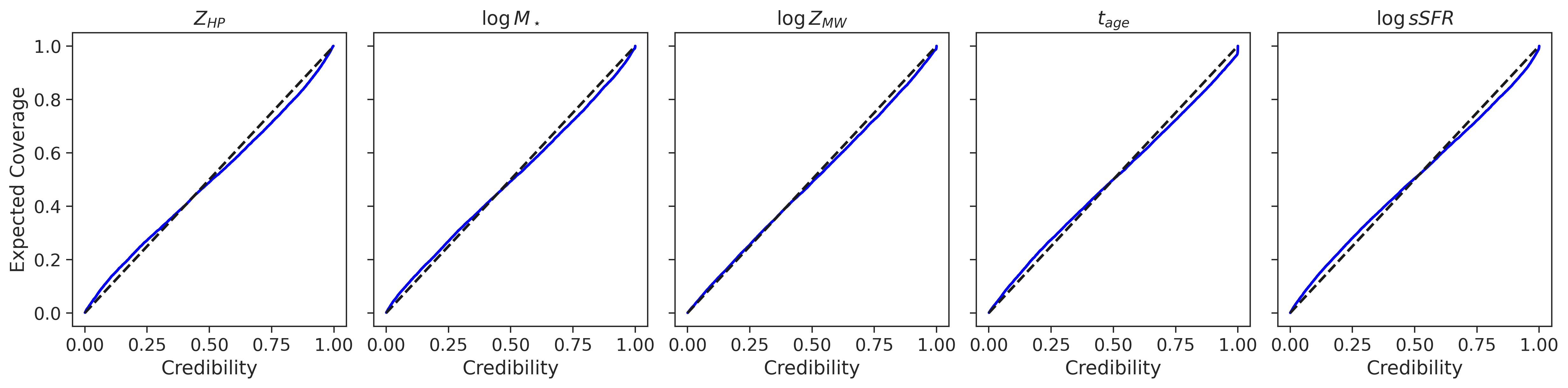}
        \caption{ResNet18 Image Embedding}
    \end{subfigure}
    
    \vspace{0.25cm} % adds vertical space between the figures
    
    \begin{subfigure}[b]{0.8\linewidth}
        \includegraphics[width=\linewidth]{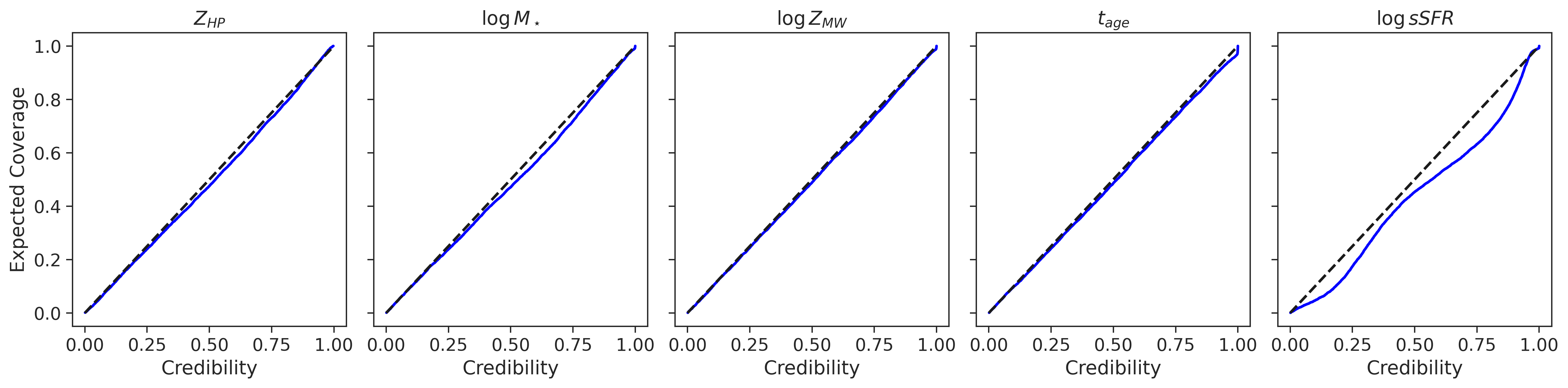}
        \caption{DINO Image Embedding}
    \end{subfigure}

    \vspace{0.25cm} % adds vertical space between the figures
    
    \begin{subfigure}[b]{0.8\linewidth}
        \includegraphics[width=\linewidth]{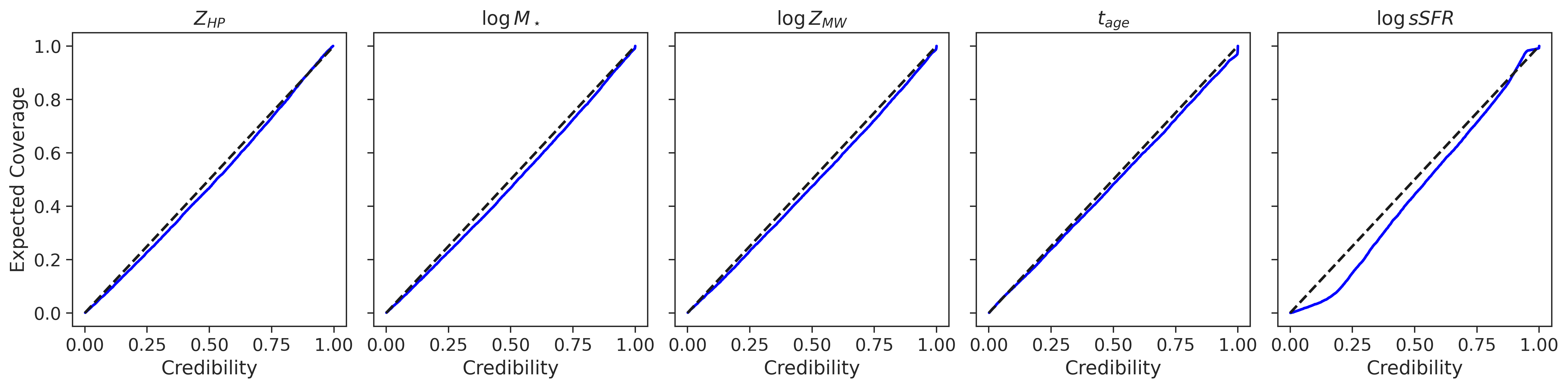}
        \caption{\cite{Stein2021} Image Embedding}
    \end{subfigure}

    \vspace{0.25cm}
    
    \begin{subfigure}[b]{0.8\linewidth}
        \includegraphics[width=\linewidth]{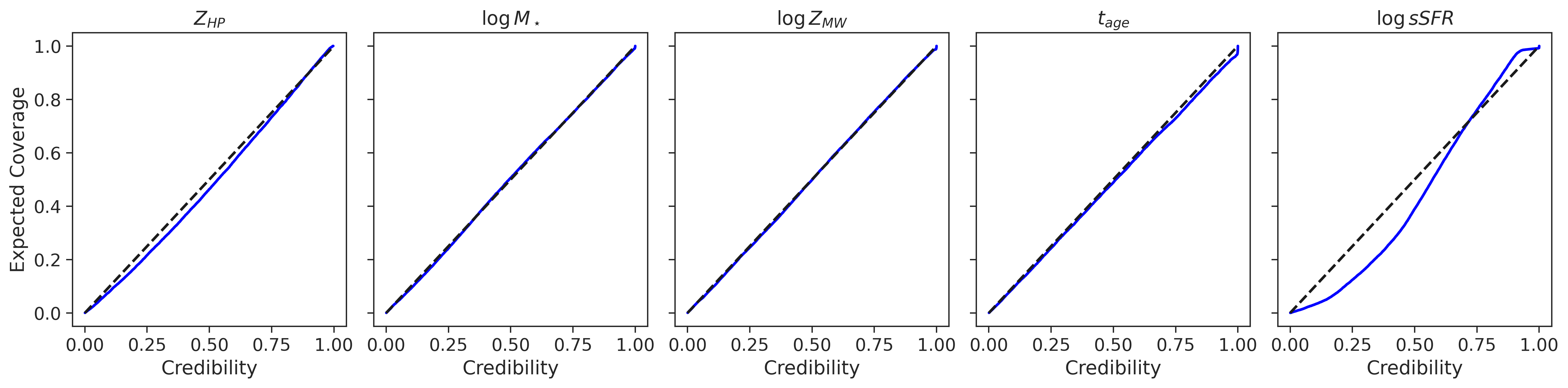}
        \caption{Photometry MLP}
    \end{subfigure}

    \caption{Tests of Accuracy with Random Points \citep[TARP;][]{lemos2023sampling} Expected Coverage Probability (ECP) tests on the trained normalizing flow ensembles for each image embedding/supervised method. If the ECP follows the diagonal line, i.e. it is equal to the confidence level for every $\alpha \in [0,1]$, then the estimator is well calibrated. Overall, the various methods appear to be well-calibrated, other than DINO and \citep{Stein2021} which are slightly biased on $\log sSFR$, and the photometry which is underconfident on $\log sSFR$.}
    \label{fig:image_flows}
\end{figure*}

\begin{figure*}
    \centering
    
    \begin{subfigure}[b]{0.8\linewidth}
        \includegraphics[width=\linewidth]{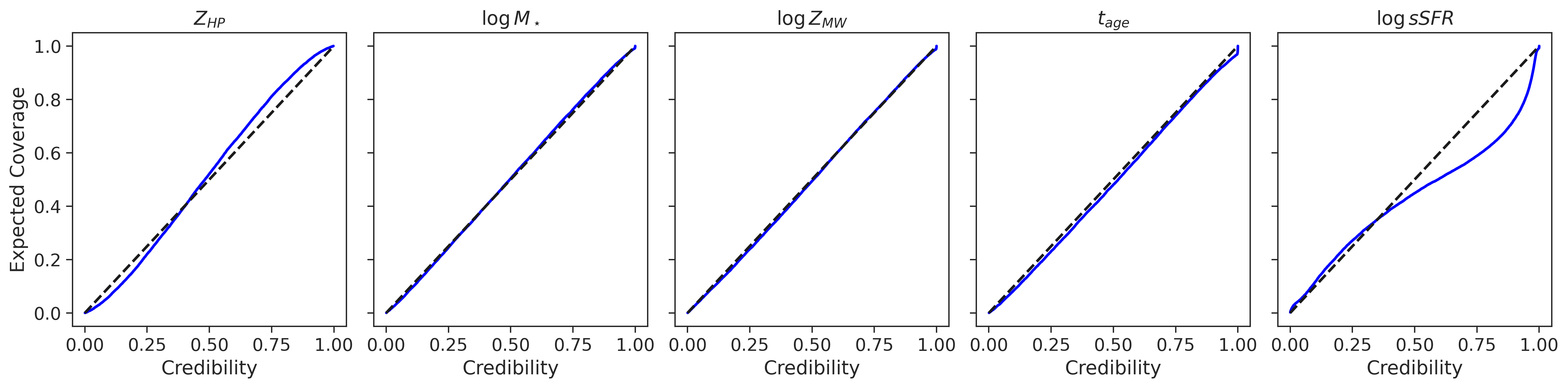}
        \caption{AstroCLIP Spectrum Embedding}
    \end{subfigure}

    \vspace{0.25cm} % adds vertical space between the figures
    
    \begin{subfigure}[b]{0.8\linewidth}
        \includegraphics[width=\linewidth]{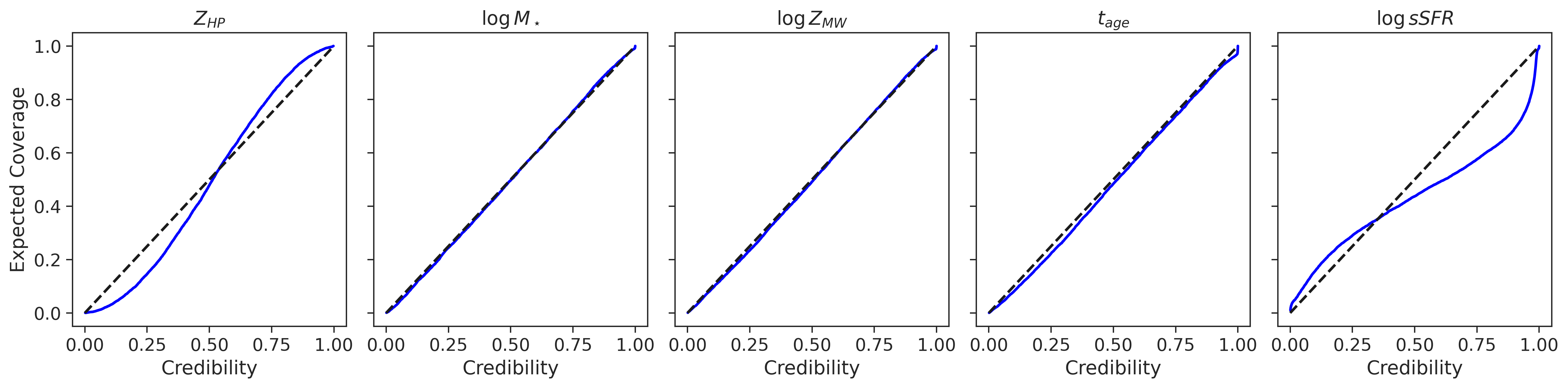}
        \caption{Spectrum Transformer Embedding}
    \end{subfigure}
    
    \vspace{0.25cm} % adds vertical space between the figures
    
    \begin{subfigure}[b]{0.8\linewidth}
        \includegraphics[width=\linewidth]{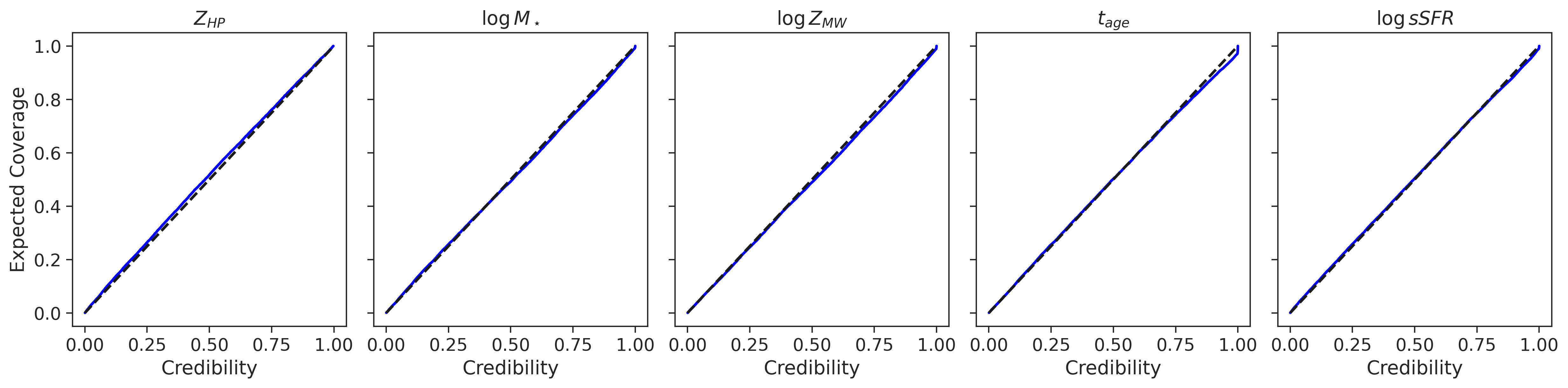}
        \caption{Spender}
    \end{subfigure}

    \caption{Tests of Accuracy with Random Points \citep[TARP;][]{lemos2023sampling} Expected Coverage Probability (ECP) tests on the trained normalizing flow ensembles for each spectrum embedding/supervised method. If the ECP follows the diagonal line, i.e. it is equal to the confidence level for every $\alpha \in [0,1]$, then the estimator is well calibrated. Overall, the various methods appear to be well-calibrated, other than CLIP and the Spectrum Transformer on $\mathcal \log sSFR$m on which they are slightly biased.}
    \label{fig:spectra_flows}
\end{figure*}

We evaluate the TARP ECP over the full dimensionality of our property space, and provide the results for the ensemble of models trained from images/photometry and from spectra in \autoref{fig:image_flows} and \autoref{fig:spectra_flows} respectively; if the ECP follows the diagonal line, i.e. it is equal to the confidence level for every $\alpha \in [0,1]$, then the estimator is well calibrated. As shown in the figures, all models are indeed well calibrated on our held-out test set on most of the property estimation tasks other than $\log sSFR$, on which some of the models are either slightly over- or under-confident.

\begin{table}
    \centering
    \caption{Galaxy morphology classification results. We train a simple MLP on the AstroCLIP galaxy image embeddings to predict the Galaxy Zoo DECaLS GZD-5 morphology classification of that galaxy. We report both the class-weighted accuracy and F1-score of the various models for each question. Overall, AstroCLIP achieves relatively strong performance on all questions, and clearly outperforms a state-of-the-art self-supervised model for galaxy images \citep{Stein2021}. We highlight in bold the best results on each question, excluding the reported ZooBot results.}
    
    \begin{subtable}{0.5\textwidth}
        \centering
        \caption{Accuracy Scores}
        \begin{tabular}{l|c|c|c|c}
        \toprule
        Question & CLIP & DINO & Stein & ZooBot \\
        \midrule
        smooth & 0.83 & \textbf{0.84} & 0.78 & 0.94 \\
        disk-edge-on & \textbf{0.97} & 0.97 & 0.87 & 0.99 \\
        spiral-arms & 0.92 & \textbf{0.95} & 0.95 & 0.93 \\
        bar & \textbf{0.56} & 0.53 & 0.53 & 0.82 \\
        bulge-size & 0.79 & \textbf{0.81} & 0.78 & 0.84 \\
        how-rounded & 0.74 & 0.79 & \textbf{0.81} & 0.93 \\
        edge-on-bulge & 0.82 & \textbf{0.86} & 0.83 & 0.91 \\
        spiral-winding & 0.74 & 0.77 & \textbf{0.79} & 0.78 \\
        spiral-arm-count & 0.44 & \textbf{0.50} & 0.50 & 0.77 \\
        merging & 0.80 & \textbf{0.81} & 0.81 & 0.88 \\
        \bottomrule
        \end{tabular}
    \end{subtable}
    
    \vspace{1cm} % Adjust the vertical space as needed
    
    \begin{subtable}{0.5\textwidth}
        \centering
        \caption{F1 Scores}
        \begin{tabular}{l|c|c|c|c}
        \toprule
        Question & CLIP & DINO & Stein & ZooBot \\
        \midrule
        smooth & 0.83 & \textbf{0.83} & 0.68 & 0.94 \\
        disk-edge-on & \textbf{0.97} & 0.97 & 0.81 & 0.99 \\
        spiral-arms & 0.94 & \textbf{0.96} & 0.95 & 0.94 \\
        bar & \textbf{0.54} & 0.37 & 0.37 & 0.81 \\
        bulge-size & 0.78 & \textbf{0.81} & 0.77 & 0.84 \\
        how-rounded & 0.74 & 0.79 & \textbf{0.81} & 0.93 \\
        edge-on-bulge & 0.81 & \textbf{0.84} & 0.75 & 0.90 \\
        spiral-winding & 0.68 & 0.73 & \textbf{0.76} & 0.79 \\
        spiral-arm-count & 0.41 & \textbf{0.47} & 0.44 & 0.76 \\
        merging & \textbf{0.73} & 0.71 & 0.71 & 0.85 \\
        \bottomrule
        \end{tabular}
    \end{subtable}
    
    \label{tab:combined_scores}
\end{table}

\subsection{Numerical Results on Galaxy Morphology Classification}
We provide the numerical results of few-shot learning from the AstroCLIP galaxy image embeddings on the Galaxy Zoo DECaLS GZD-5 survey detailed in \autoref{sec:galaxy_zoo}. We only evaluate galaxy classes on galaxies for which more than 50\% of the volunteers shown that galaxy answered that question.

% Don't change these lines
\bsp	% typesetting comment
\label{lastpage}
\end{document}